\documentclass[12pt]{article}
\usepackage{graphicx}
 \oddsidemargin  = - 7 mm 
 \evensidemargin = - 7 mm
\def \ketv #1>{\mbox{$|{#1}\rangle$}} 
\def \etal{{\it et al.}}
\newcommand{\rts}{ \sqrt s}
\begin{document}
\date{}
\title{Chiral unitary approach to S-wave meson baryon scattering \\
       in the strangeness $S=0$ sector} 
\author{
 {\large T.~Inoue\thanks{e-mail: inoue@ific.uv.es},
    E.~Oset and M.J. Vicente Vacas} \\
 {\small Departamento de F\'{\i}sica Te\'orica and IFIC, 
  Centro Mixto Universidad de Valencia-CSIC }\\
 {\small Institutos de Investigaci\'on de Paterna,
  Apdo. correos 22085, 46071, Valencia, Spain }
       } 
\maketitle
\begin{abstract}
 We study the S-wave interaction of mesons with baryons
 in the strangeness $S=0$ sector in a coupled channel unitary approach.
 The basic dynamics is drawn from the lowest order meson baryon
 chiral Lagrangians.
 Small modifications inspired by models with explicit vector meson exchange
 in the $t-$channel are also considered.
 In addition the $\pi \pi N$ channel
 is included and shown to have an important repercussion in the results,
 particularly in the $I=3/2$ sector. 
 The $N^*(1535)$ resonance is dynamically generated and appears
 as a pole in the second Riemann sheet with its mass,
 width and branching ratios in fair agreement with experiment.
 A $\Delta(1620)$ resonance also appears as a pole 
 at the right position although with a very large width,
 coming essentially from the coupling to the $\pi \pi N$ channel,
 in qualitative agreement with experiment.
\end{abstract}

\section{Introduction}

 The introduction of effective chiral Lagrangians to account for the basic
 symmetries of QCD and its application through $\chi PT$ to the study of
 meson meson interaction \cite{gasser} or meson baryon interaction 
 \cite{ulf,veronique,pich,ecker} has brought new light into these
 problems and allowed a systematic approach. 
 Yet, $\chi PT$ is constrained to the low energy region, 
 where it has had a remarkable success, 
 but makes unaffordable the study of the intermediate energy region 
 where resonances appear. 
 In recent years, however, 
 the combination of the information of the chiral Lagrangians,
 together with the use of nonperturbative schemes, have allowed one
 to make prediction beyond those of the chiral perturbation expansion.
 The main idea that has allowed the extension of $\chi PT$ to higher energies
 is the inclusion of unitarity in coupled channels.
 Within the framework of chiral dynamics, 
 the combination of unitarity in coupled channels together with a reordering 
 of the chiral expansion, provides a faster convergence
 and a larger convergence radius of a new chiral expansion, 
 such that the lowest energy resonances are generated within those schemes. 
 A pioneering work along this direction was made in \cite{siegel,wolfram,kaiser}
 where the Lippmann Schwinger equation in coupled channels was used
 to deal with the meson baryon interaction in the region
 of the $N^*(1535)$ and $\Lambda(1405)$ resonances. 
 Similar lines, using the Bethe Salpeter equation in the meson meson interaction,
 were followed in \cite{npa} and a more elaborated framework was subsequently
 developed using the Inverse amplitude method (IAM) \cite{ramonet} and the N/D
 method \cite{nsd}. 
 The IAM method was also extended to the case of the meson baryon interaction
 in \cite{nicola} and in \cite{nicolajuan} where a good reproduction 
 of the $\Delta$ resonance was obtained using second order parameters of natural size.
 The N/D method has also been used for $\pi N$
 scattering \cite{joseulfpi} and in the $K^- N$ and coupled channels system in
 \cite{joseulflan}. 
 A review of these unitary methods can be seen in \cite{report}. 
 The consideration of coupled channels to study meson baryon interactions at
 intermediate energies has also been exploited in \cite{dytman} 
 using the K-matrix approach, although not within a chiral context.
  
 The Bethe Salpeter equation was also used in the  study of
 the the meson baryon interaction in the strangeness $S=-1$ sector in \cite{angels}
 and in the $S=0$ sector around the $N^*(1535)$ region in \cite{hosaka}. 
 In this latter work, aimed at determining the $N^*N^* \pi$ coupling,
 only the vicinity of the resonance was studied and no particular attention
 was given to the region of lower energies.
 Subsequently a work along the same lines using the Bethe Salpeter equation, 
 but considering all the freedom of the chiral constraints, was done in \cite{juan}
 and a good reproduction of the experimental 
 observables was obtained for the $I=1/2$ sector. 
 Those works considered only states of meson baryon in the coupled channels
 and both in \cite{hosaka} and \cite{juan} the $\pi \pi N$ channel was omitted.
 This channel plays a moderate role in the $I=1/2$ sector \cite{wycech} but,
 as we shall see, it plays a crucial role in the $I=3/2$ sector. 
  
 Our aim in the present work is to extend the chiral unitary approach to
 account for the $\pi \pi N$ channel, including simultaneously some other
 corrections inspired by vector meson dominance (VMD) which finally allow 
 one to have a reasonable description of the meson baryon interaction
 up to meson baryon energies of around 1600 MeV. 
 The $N^*(1535)$ resonance is generated dynamically in this approach and the mass,
 width and branching ratios are obtained in fair agreement with the experiment. 
 The phase shifts and inelasticities for $\pi N$ scattering in that
 region are also evaluated and good agreement with experiment is also found
 both in the $I=1/2$ and $I=3/2$ sectors. 
 In addition some trace of the $\Delta(1620)$ is found, 
 linked to the introduction of the $\pi \pi N$ channel,
 with a pole in the second Riemann sheet with the right energy, albeit a large width.

 A side effect of the calculations is that we determine the s-wave part of
 the $\pi N \to \pi \pi N $ transition amplitude, revising previous 
 determinations in \cite{manley} and \cite{burkhardt}, 
 and together with the P-wave amplitudes previously determined, 
 we obtain a good reproduction of the cross sections for these reactions.

\section{$\pi N$ scattering in a 2-body coupled channel model}
 
\subsection{Basic 2-body model}

\begin{figure}[t]
\begin{center}
 \includegraphics[height=13mm]{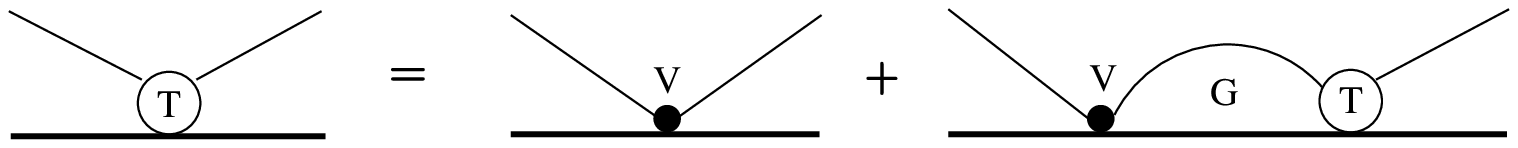}
 \caption{ Diagram representation of the Bethe-Salpeter equation. }
 \label{fig:bseq}
\end{center}
\end{figure}
 
 In this section we study the meson baryon scattering in S-wave in the
 strangeness $S=0$ sector. We shall make use of the Bethe Salpeter equation in
 coupled channels considering states of a meson of the $0^-$ octet and a
 baryon of the $1/2^+$ octet, as required by the SU(3) chiral formalism. 
 For total zero charge we have six channels,
 $\pi^- p$, $\pi^0 n$, $\eta n$, $K^+ \Sigma^-$,
 $K^0 \Sigma^0$, and $K^0 \Lambda$. 
 
 The Bethe Salpeter equation is given by 
\begin{equation}
  T = V + V G T
\label{eqn:bseq}
\end{equation}
 where $G$ is the product of the meson and baryon propagators. 
 The diagrammatic expression of the $T$ matrix is shown in Fig.\ref{fig:bseq}.
 Following \cite{angels} we take the kernel (potential) of
 the Bethe Salpeter equation from the lowest order chiral Lagrangian involving
 mesons and baryons 
\begin{equation}
 {\cal L}_{MB \to M'B'}
 = Tr \left[ \bar B i \gamma^{\mu} \frac{1}{4 f^2} 
    \left\{    (\phi \partial_{\mu} \phi - \partial_{\mu} \phi  \phi) B
          - B (\phi \partial_{\mu} \phi - \partial_{\mu} \phi  \phi) 
    \right\}
       \right]
\label{eqn:lagrangian}
\end{equation}
 where $B$ and $\phi$ are the SU(3) matrices for the octet baryon field and
 the octet meson field respectively, and $f$ is the weak decay constant of the meson.
 From this Lagrangian, the transition potentials 
 between our six channels are given by 
\begin{equation}
 V_{ij} = - C_{ij}\frac{1}{4f^2}\bar u(p')\gamma^{\mu} u(p) (k_{\mu}+k'_{\mu} )
 \label{eqn:chv}
\end{equation}
 with the initial(final) baryon spinor $u(p)$ ($u(p')$) and 
 the initial(final) meson momentum $k$ ($k'$).
 The coefficients  $C_{ij}$, reflecting the SU(3) symmetry of the problem,
 are obtained from eq. (\ref{eqn:lagrangian}) and shown in Table \ref{tbl:cij}.

\newcommand{\chone}{$K^+ \Sigma^-$}
\newcommand{\chtwo}{$K^0 \Sigma^0$}
\newcommand{\chthr}{$K^0 \Lambda$}
\newcommand{\chfou}{$\pi^-  p$}
\newcommand{\chfiv}{$\pi^0  n$}
\newcommand{\chsix}{$\eta   n$}
\newcommand{\chsev}{$\pi^0 \pi^- p$}
\newcommand{\cheig}{$\pi^+ \pi^- n$}

\begin{table}[t]
\caption{$C_{ij}$ coefficient in the potential. $C_{ij}=C_{ji}$} 
\begin{center}
\begin{tabular}{ccccccc}
 \hline
        & \chone &  \chtwo & \chthr & \chfou        & \chfiv            & \chsix
 \\
 \hline
 \chone &  1  & $-\sqrt2$ & 0   & 0                 & $-\frac1{\sqrt2}$ & $-\sqrt{\frac32}$ 
 \\
 \chtwo &     &  0        & 0   & $-\frac1{\sqrt2}$ & $-\frac12$        & $\frac{\sqrt3}2$
 \\
 \chthr &     &           & 0   & $-\sqrt{\frac32}$ & $\frac{\sqrt3}2$  & $-\frac32$
 \\
 \chfou &     &           &     & 1                 & $-\sqrt2$         & 0
 \\
 \chfiv &     &           &     &                   & 0                 & 0
 \\
 \chsix &     &           &     &                   &                   & 0
 \\
 \hline
\end{tabular}
\end{center}
\label{tbl:cij}
\end{table}

 We are interested in the value of the T matrix element
 for the on-shell meson-baryon system 
 at a certain center of mass energy $\rts$
 or $P \equiv p+k = (\rts,0,0,0)$, 
 and we express it as $T(\rts)$.
 Yet, in the diagrammatic expression of the Bethe Salpeter equation, 
 Fig.\ref{fig:bseq}, one can see that the $V$ and $T$ matrices
 in the second diagram on the right hand side,
 can be (half) off-shell since the BS equation is an integral equation.
 However, it was shown in \cite{angels} that the off shell part of these
 matrices in the loops could be incorporated into a renormalization of the
 lowest order Lagrangian and hence only the on shell parts are needed. 
 Thus, they factorize out of the integral and the original BS integral equation
 is then reduced to an algebraic equation which is easily solved with the 
 solution in matrix form
\begin{equation}
   T(\rts)  = [V(\rts)^{-1} - G(\rts)]^{-1}
   ~~~~\mbox{or}~~~~
   T(\rts)^{-1} = V(\rts)^{-1} - G(\rts)
   \label{eqn:sol}
\end{equation}
 where the $G$ function is a diagonal matrix representing the loop
 integral of a meson and a baryon. 
 The $i$-th element is thus expressed as  
\begin{equation}
  G_i(P) = i
           \int \frac{d^4 q}{(2 \pi)^4} 
           \frac{2 M_i}{(P-q)^2-M_i^2 + i \epsilon}
           \frac{1}{q^2 - m_i^2 + i \epsilon } 
\label{eqn:oneloop}
\end{equation}
 with $M_i$ the baryon mass and $m_i$ the meson mass.

 In the present (Mandl and Shaw) normalization \cite{mandl},
 the T matrix is related to the S matrix as
\begin{equation}
  S_{ij}(P_i,P_j)
  = 
  1 - i (2 \pi)^4 \delta^4(P_i - P_j) \frac{1}{V^2}
        \sqrt{ \frac{M_i}{E_i} }  \sqrt{ \frac{M_j}{E_j} }
        \sqrt{ \frac{1}{2 \omega_i} } \sqrt{ \frac{1}{2 \omega_j} } 
        {\cal T}_{ij}(P_i,P_j)
\end{equation}
 where V is the volume of the normalization box, 
 $E_i$ ($E_j$) and $\omega_i$ ($\omega_j$) are the 
 energies of the incoming (outgoing) baryon and meson respectively.
 In this notation, the unitarity condition for any partial wave amplitude
 (in particular the S-wave which is our case)  
 is given by
\begin{equation}
 \mbox{Im}[T_{ij}(\rts)] = 
    - T_{ik}(\rts) \frac{ M_k Q_k(\rts)}{4 \pi \rts} T_{kj}^*(\rts)
   ~~~~\mbox{or}~~~~
 \mbox{Im}[ T^{-1}(\rts)]_{ij} = \delta_{ij} \frac{ M_i Q_i(\rts)}{4 \pi \rts}
 \label{eqn:unitarity}
\end{equation}
 where 
\begin{equation}
 Q_i(\rts) = \frac{ \sqrt{ \left(s - (M_i+m_i)^2 \right)
                      \left(s - (M_i-m_i)^2 \right) } }{2 \rts }
             \theta( \rts - (M_i + m_i) )
\end{equation}
  is the on shell center of mass momentum of $i$-th meson-baryon system.

 The matrix elements of the potential which we substitute into eq. (\ref{eqn:sol}) is  
\begin{equation}
  V_{ij}(\rts) = - C_{ij}\frac{1}{4 f_i f_j} (2 \rts - M_i - M_j)
           \sqrt{ \frac{M_i + E_i(\rts) }{2 M_i} }
           \sqrt{ \frac{M_j + E_j(\rts) }{2 M_j} }
   \label{eqn:pot}
\end{equation}
 where we introduce different weak decay constants for different mesons.
 We use the values 
\begin{equation}
 f_{\pi}=93 ~\mbox{MeV} ~,~ f_K = 1.22 f_{\pi} ~,~ f_{\eta}=1.3 f_{\pi}
\end{equation}
 taken from $\chi$PT \cite{gasser}.

 The meson-baryon 2-body propagator 
 which we substitute into  equation (\ref{eqn:sol}) is  
\begin{eqnarray}
 \mbox G_i(\rts) &=&  \frac{2 M_i}{(4 \pi)^2} 
  \left\{ 
        a_i(\mu) + \log \frac{m_i^2}{\mu^2} + 
        \frac{M_i^2 - m_i^2 + s}{2s} \log \frac{M_i^2}{m_i^2} 
  \right.
  \\
     &+& \frac{Q_i(\rts)}{\rts} 
    \left[
         \log \left(  s-(M_i^2-m_i^2) + 2 \rts Q_i(\rts) \right) 
      +  \log \left(  s+(M_i^2-m_i^2) + 2 \rts Q_i(\rts) \right) 
    \right.
  \nonumber
  \\
  & & 
  \Biggl.
    \left.
      - \log \left( -s+(M_i^2-m_i^2) + 2 \rts Q_i(\rts) \right) 
      - \log \left( -s-(M_i^2-m_i^2) + 2 \rts Q_i(\rts) \right)
    \right]
  \Biggr\}  
  \nonumber
\end{eqnarray}
 which is the 1-loop integral (\ref{eqn:oneloop}) done with dimensional regularization.
 Its infinity is canceled by higher order counter-terms.
 The first terms $a_i(\mu)$ are real constants and stand for the finite 
 contribution of such counter-terms. 
 We treat these $a_i(\mu)$ as unknown parameters and determine them
 from fits to the data.
 The imaginary part of the above propagator is 
\begin{equation}
  \mbox{Im}[ G_i(\rts) ] = - \frac{M_i Q_i(\rts)}{4 \pi \rts} 
  \label{eqn:imgg}
\end{equation}
 which leads, using eq. (\ref{eqn:sol}), to eq. (\ref{eqn:unitarity}).
 Therefore, in the present model, unitary is exactly fulfilled.

\begin{figure}[t]
\begin{center}
    \includegraphics[height=80mm]{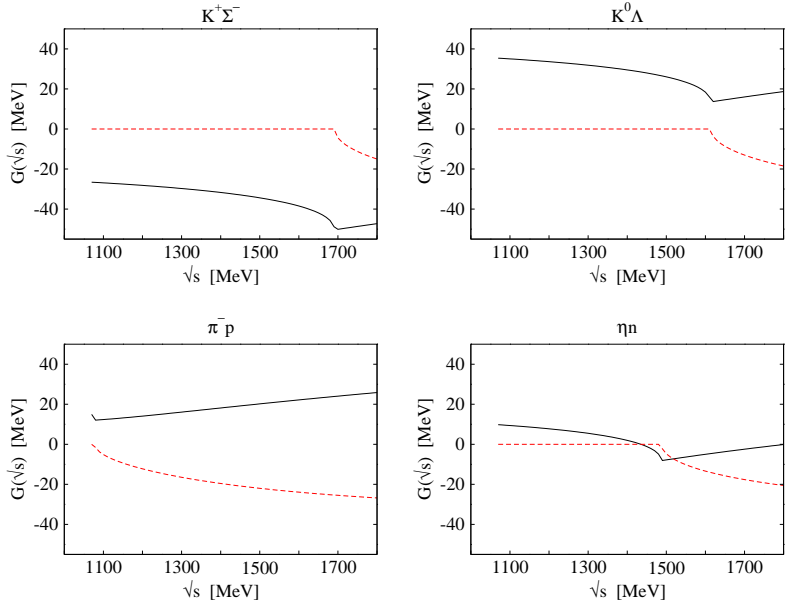}
\end{center}
  \caption{ 
   Propagators of the meson-baryon systems
   using the parameters of eq (\ref{eqn:subjose}).
   The $K^0 \Sigma^0$ and $\pi^0 n$ propagators are omitted 
   because they are almost the same as those of $K^+ \Sigma^-$ and $\pi^- p$ 
   respectively.
   The continuous(dashed) lines stand for the real(imaginary) part.
          }
  \label{fig:propjose}
\end{figure}

 In order to keep isospin symmetry in the case that the masses of the particles
 in the same multiplet are equal, we choose $a_i(\mu)$ to be the same for the states
 belonging to the same isospin multiplet. 
 Hence we have four subtraction constants
 $a_{\pi N}(\mu), a_{\eta N}(\mu), a_{K \Lambda}(\mu)$ and $a_{K \Sigma}(\mu)$.
 A best fit to the data with eq. (\ref{eqn:sol}) leads us to the
 following values of these parameters
\begin{equation}
   \mu= 1200 ~\mbox{MeV},~~ 
   a_{\pi N}(\mu)     =  2.0 ,~~
   a_{\eta N}(\mu)    =  0.2 ,~~
   a_{K \Lambda}(\mu) =  1.6 ,~~ 
   a_{K \Sigma}(\mu)  = -2.8 ~~.
\label{eqn:subjose}
\end{equation} 
 In Fig.\ref{fig:propjose}, we show the $G$ matrix elements which are obtained
 with this set of parameters.
 These propagators are essentially the same as those 
 obtained in \cite{hosaka} where the regularization 
 was done using a three-momentum cutoff ($|\vec q| < 1 ~\mbox{GeV}$)
\begin{equation}
  \mbox{Re}[G_i(\rts)] = 
     {\cal P} \int^{1 GeV} \! \! \!  \frac{d^3 \vec q}{(2 \pi)^3}
     \frac{M_i}{E(\vec q)} 
     \frac{1}{\rts - E(\vec q) - \omega(\vec q)}
     \frac{1}{2 \omega(\vec q)} 
     + \tilde a_i
\end{equation}
 with $E(\vec q)= \sqrt{M_i^2 + \vec q\,^2}$ , 
      $\omega(\vec q) = \sqrt{ m_i^2 + \vec q\,^2}$ and 
\begin{equation}
  \tilde a_{\pi N}     = 35 ~\mbox{MeV} ,~
  \tilde a_{\eta N}    = 16 ~\mbox{MeV} ,~
  \tilde a_{K \Lambda} = 40 ~\mbox{MeV} ,~
  \tilde a_{K \Sigma}  =-21 ~\mbox{MeV}
\end{equation}
 where ${\cal P}$ means the Cauchy principal value. 

\begin{figure}[t]
 \begin{center}
  \includegraphics[height=80mm]{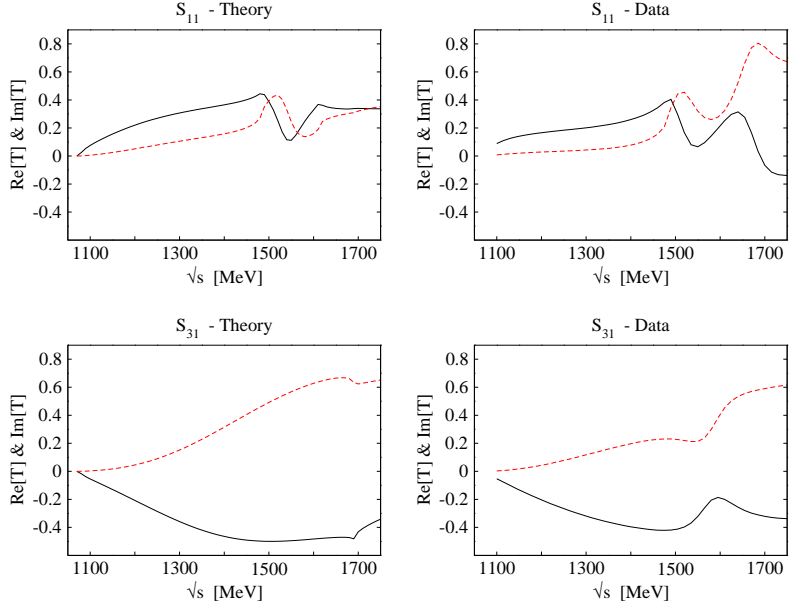}
 \end{center}
 \caption{
   Scattering amplitude for the $S_{11}$ and $S_{31}$ $\pi N$ partial waves. 
   The left and right figures stand for the amplitudes calculated in our model
   and the data analysis \cite{CNS} respectively.
   The continuous(dashed) lines stand for the real(imaginary) part.
           }
  \label{fig:tmatjose}
\end{figure}

 The resulting T matrix elements of $S_{11}$ (isospin 1/2) and 
 $S_{31}$ (isospin 3/2) $\pi N$ elastic scattering 
 are shown in Fig.\ref{fig:tmatjose}.
 In these graphs we plot the quantity
\begin{equation}
    -  \sqrt{ \frac{M_i Q_i(\rts)}{4 \pi \rts} } 
       \sqrt{ \frac{M_j Q_j(\rts)}{4 \pi \rts} } T_{ij}(\rts)
\end{equation}
 in order to compare with the data of the CNS analysis \cite{CNS}.
 We find a qualitative agreement with the data in  
 the energy range from threshold to 1600 MeV.
 In the figure of the experimental analysis one can see the manifestation of  
 the resonances  $N^*(1535)$, $N^*(1650)$ in the $S_{11}$ amplitude 
 and $\Delta(1620)$ in the $S_{31}$  one.
 The calculated $S_{11}$ amplitude also exhibits a resonance 
 structure around 1535 MeV. 
 The generation of this resonance is common to
 all the unitary chiral approaches \cite{siegel,hosaka,juan}. 
 In section 4, after we include new elements in the theory, 
 we shall investigate this resonance by
 searching for poles in the second Riemann sheet of the complex plane.
 
 At energies beyond 1600 MeV, 
 the calculated amplitudes are qualitatively different from the data,
 and the $N^*(1650)$ and $\Delta(1620)$ do not show up. 
 In the philosophy that the resonances obtained using the lowest order Lagrangian
 and the present unitary scheme are simply meson baryon scattering resonances, 
 (qualifying for quasibound meson baryon states \cite{report}), 
 the nongeneration of a particular resonance would indicate 
 that it is mostly a genuine state (approximately a 3q system). 
 However, such resonances could be also obtained 
 in a unitary approach provided one used information related to this resonance
 which would be incorporated in the higher order Lagrangians.
 For the case of the meson meson interaction it is known \cite{derafael} 
 that the $O(p^4)$ Lagrangian is saturated by the exchange of vector meson resonances,
 which are not generated in the BS approach of \cite{npa}. 
 Actually, in \cite{juan} the $N^*(1650)$ resonance is also
 reproduced by introducing counter terms which effectively account for higher
 order corrections, much in the way as the (genuine) $\rho$ resonance was
 reproduced in the study of the meson meson  scattering in \cite{bsjuan1,bsjuan2}.
 As quoted before, the agreement below 1600 MeV is only qualitative. 
 Indeed, both the real and imaginary parts of the $S_{11}$ amplitude are somewhat
 overestimated in the theory, and so is the case for the $S_{31}$ amplitude
 where the theoretical imaginary part clearly overestimates the experimental one. 

\begin{figure}[t]
\begin{center}
 \includegraphics[height=100mm]{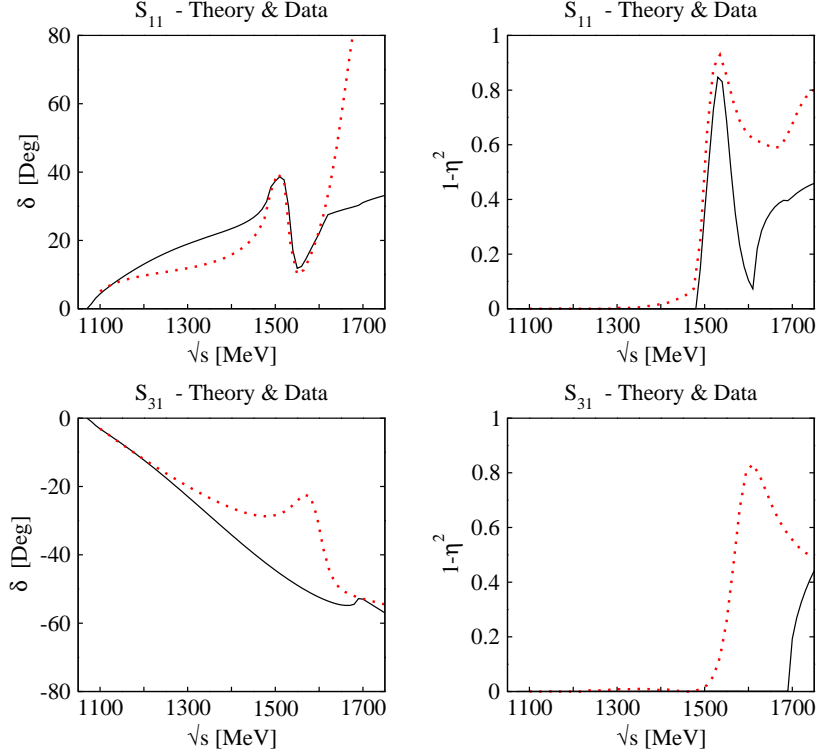}
 \caption{ 
   Phase-shifts and inelasticities for $S_{11}$ and $S_{31}$ $\pi N$ scattering.
   The continuous(dotted) lines correspond the calculations(data analysis \cite{CNS}).
           }
 \label{fig:phinjose}
\end{center}
\end{figure}

 The phase-shifts $\delta_i(\rts)$ and elasticities $\eta_i(\rts)$ are given by
\begin{eqnarray}
 \delta_i(\rts) 
   &=& 
   \frac12 \arccos\left[ 
       \mbox{Re}\left[
   \frac{ 1 - 2 i  \frac{M_i Q_i}{4 \pi \rts} T_{ii}   }
        {\left| 1 - 2 i  \frac{M_i Q_i}{4 \pi \rts} T_{ii} \right|}
                \right] 
                 \right]
   \times
   \mbox{sign}\left[ 
      \mbox{Re}\left[- T_{ii} \right]
              \right]
 \\
 \eta_i(\rts)  
   &=& \left| 1 - 2 i  \frac{M_i Q_i}{4 \pi \rts} T_{ii} \right| ~~.
\end{eqnarray}
 Fig.\ref{fig:phinjose} shows the phase-shifts and inelasticities. 
 In the phase-shifts graph we can see again the qualitative agreement
 at  energies below 1600 MeV.
 On the other hand, the inelasticities are not reproduced
 below the first open meson baryon threshold.
 In this 2-body model 
 the threshold of inelastic scattering for the  $S_{11}$ case  
 is the $\eta n$ threshold which appears at 1487 MeV. 
 Below this energy, the calculated inelasticities are zero and do not agree 
 with the data.
 This situation is even clearer in the  $S_{31}$ case.
 The threshold of inelastic scattering is in this case the $K \Sigma$ threshold
 which appears at 1690 MeV.
 The big inelasticities observed in the data below that energy are not 
 reproduced in the present 2-body approach. The only inelastic channel opened
 below that energy is the $\pi \pi N$ channel, and the experimental data is
 telling us that the influence of this channel in the $S_{31}$ amplitude should
 be very important.
 We will include the $\pi \pi N$ channel in the next section.

\subsection{Improved 2-body model}

\begin{figure}[t]
\begin{center}
 \includegraphics[height=40mm]{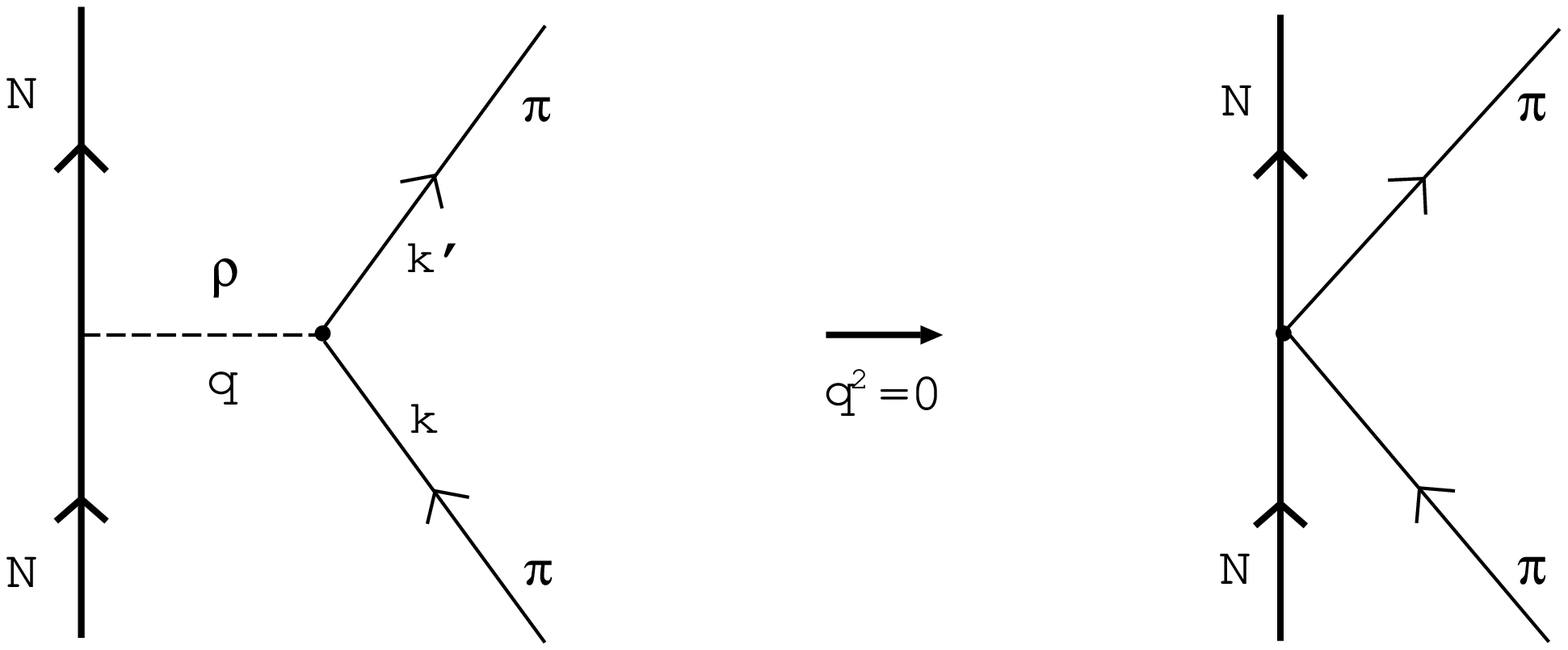} 
 \caption{The mechanism of $\pi N \to \pi N$ interaction
          in the vector meson dominance hypothesis.}
 \label{fig:rhoex}
\end{center}
\end{figure}

\begin{figure}[t]
\begin{center}
 \includegraphics[height=40mm]{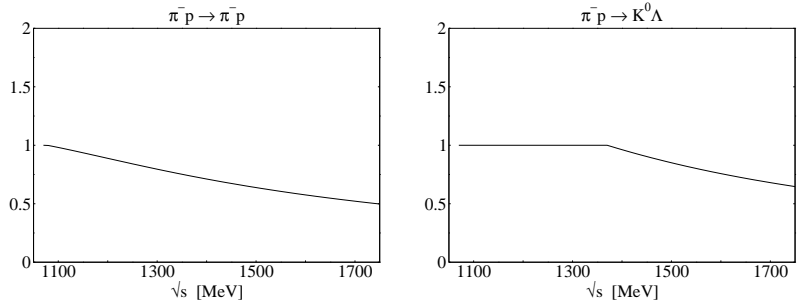}
 \caption{Correction factor of the chiral coefficient.}
 \label{fig:correc}
\end{center}
\end{figure}

 The vector meson dominance(VMD) hypothesis is phenomenologically very successful.
 In this hypothesis, the meson baryon interaction is provided
 by vector meson exchange.
 For example, the $\pi N \to \pi N$ process is described by 
 $\rho$ meson exchange in the t-channel, 
 as shown in Fig.\ref{fig:rhoex}.
 It is interesting to note that the result from $\rho$ exchange provides the
 amplitude obtained from the lowest order chiral Lagrangian.
 For example, in the case of $\pi^- p \to \pi^- p$, 
 $\rho$ exchange in the $t-$channel gives
 (see \cite{wambach} for the $\rho NN $ coupling within the VMD hypothesis) 
\begin{equation}
 \left.
 i \frac{m_{v}G_{v}}{2 f^2} \gamma^{\mu} \epsilon_{\mu}
 \frac{i}{q^2 - m_{v}^2}
 i \frac{m_{v}G_{v}}{f^2} \epsilon^{\nu} (k+k')_{\nu}
 \right|_{q^2=0}
 =  -i \frac{1}{4f^2}\gamma^{\mu}(k+k')_{\mu}
 \label{eqn:ampinvmd}
\end{equation}
 which follows from the relation $G_{v}^2/f^2=1/2$,
 which is a result of the VMD hypothesis.
 This indicates that the lowest order chiral Lagrangian is an
 effective manifestation of the VMD mechanism in the vector
 field representation of the vector mesons. According to this consideration, 
 we introduce in the chiral coefficient a correction to account for
 the dependence on the momentum transfer of the $\rho$ propagator. 
 Thus we replace  
\begin{equation}
  C_{ij} \ \rightarrow \
  C_{ij} \times \int
           \! \frac{d \hat k'}{4 \pi} \, \frac{-m_{v}^2}{(k'-k)^2-m_{v}^2} 
  ~~~\mbox{at}~~~ \rts > \sqrt{s_{ij}^{0}}
 \label{eqn:modify}
\end{equation}
 where $\sqrt{s_{ij}^{0}}$ is the energy where the integral 
 of eq. (\ref{eqn:modify}) is unity, 
 and which appears in between the thresholds of the two $i,j$ channels. 
 At very low energies where $\chi PT$ is used, this correction is negligible
 but this is not the case at the intermediate energies studied here.
 For example, the correction for $\pi^- p \to \pi^- p$ element is calculated as
\begin{equation}
  \int \! \frac{d \hat k'}{4 \pi} \, \frac{-m_{\rho}^2}{(k'-k)^2-m_{\rho}^2} 
  =
  \frac{m_{\rho}^2}{4 k k'}
   \log \frac
   { m_{\rho}^2 + 2 k^0 k'^0 + 2 k k' - m_{\pi}^2 - m_{\pi}^2} 
   { m_{\rho}^2 + 2 k^0 k'^0 - 2 k k' - m_{\pi}^2 - m_{\pi}^2}
\end{equation}
 with $m_{\rho}=770 \mbox{[MeV]}$,
 and is shown in Fig.\ref{fig:correc} left.
 One can see that $\rho$ meson tail reduces the coefficient about 25\%
 at  energies around 1500 MeV.
 Similarly, for the strangeness exchanging process,
 we consider $K^*$ exchange in the t-channel.
 For example, the correction of $\pi^- p \to K^0 \Lambda$ 
 is calculated with $m_{K^*}=892$ MeV 
 and shown in Fig.\ref{fig:correc} right.

\begin{figure}[t]
\begin{center}
   \includegraphics[height=40mm]{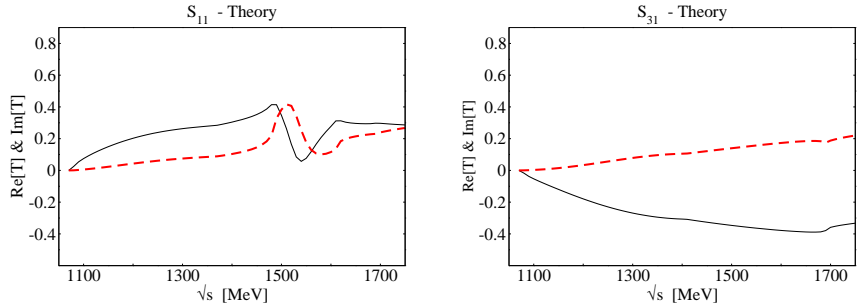}
 \caption{ 
   Scattering amplitude for the $S_{11}$ and $S_{31}$ $\pi N$ partial waves 
   with the improved $C_{ij}$.
         } 
 \label{fig:tmatvmd}
\end{center}
\end{figure}

\begin{figure}[t]
\begin{center}
  \includegraphics[height=50mm]{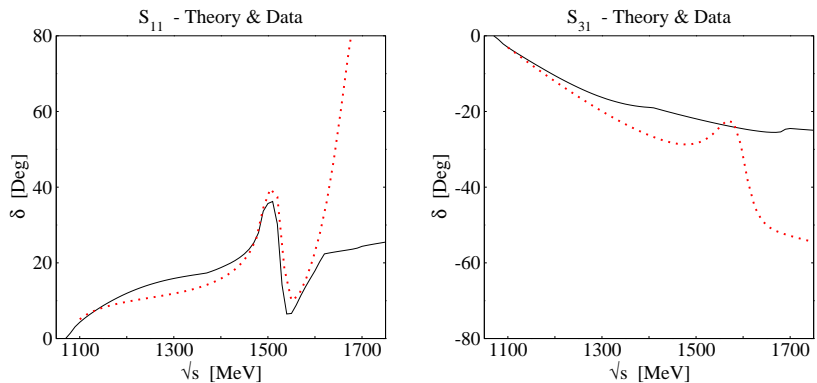}
  \caption{ 
   Phase-shifts for $S_{11}$ and $S_{31}$ $\pi N$ scattering 
   with the improved $C_{ij}$.
          } 
\label{fig:phvmd}
\end{center}
\end{figure}

 With this correction, after retuning the subtraction constants $a_i(\mu)$,
 we obtain the T matrix shown in Fig.\ref{fig:tmatvmd}.
 As we can see, the problem of the previous overestimation is nearly solved. 
 Especially, a drastic improvement is achieved for the $S_{31}$ amplitude. 
 The phase-shifts are better reproduced now as shown in Fig.\ref{fig:phvmd}. 
 These results show that the correction of eq. (\ref{eqn:modify}) 
 is important and leads to improved results with respect to the 
 plain use of the standard lowest order chiral Lagrangian.
 In the following sections, we employ this modified coefficient. 
 The calculated inelasticities with this coefficient,
 are almost the same as before and show the lack of some important channels. 
 We come back to this problem in the next section.

\section{$\pi N$ scattering in a 2-3-body coupled channel model}

  In this section, we extend our model to include the $\pi \pi N$ channels.
  The cross sections of the $\pi N \to \pi \pi N$ scattering 
  are known experimentally and they are sizeable compared with the two body
  cross sections. 
  In this paper, 
  we include only the transition potential between $\pi N$ and $\pi \pi N$
  and disregard the coupling  between the 
  $\{K \Sigma, K \Lambda, \eta n, \pi \pi N\}$ states and the $\pi \pi N$.
  This is a simplification forced by the ignorance of such couplings, 
  but the larger mass of these states with respect to $\pi N$, 
  and the fact that we are talking about corrections,
  make this simplification justifiable.

\subsection{$\pi N \to \pi \pi N$ process}

\begin{figure}[t]
\begin{center}
 \includegraphics[height=25mm]{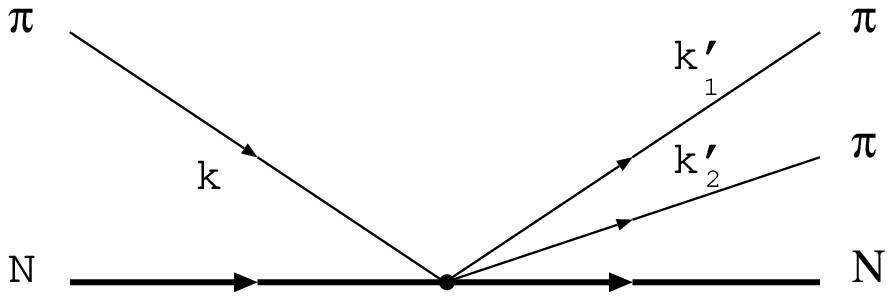}
 \caption{$\pi N \to \pi \pi N$ process}
 \label{fig:twopi}
\end{center}
\end{figure}

  In the isospin formalism, 
  the $\pi N \to \pi \pi N$ transition amplitudes can be classified by 
  the total isospin $I$ and the isospin of the final two pions $I_{\pi\pi}$,
  and the corresponding amplitudes are written as $A_{2I I_{\pi\pi}}$.
  The amplitudes of the physical processes
  are expressed in terms of the following four independent isospin amplitudes 
\begin{equation}
\begin{array}{lll}
   A_{11}=a_{11}\,\chi_f^{\dagger}\vec\sigma\!\cdot\!(\vec k'_1-\vec k'_2) \chi_i
   &~,~ &
   A_{31}=a_{31}\,\chi_f^{\dagger}\vec\sigma\!\cdot\!(\vec k'_1-\vec k'_2) \chi_i
   \\
   A_{10}=a_{10}\,\chi_f^{\dagger}\vec\sigma\!\cdot\!\vec k \chi_i
   &~,~&
   A_{32}=a_{32}\,\chi_f^{\dagger}\vec\sigma\!\cdot\!\vec k \chi_i
\end{array}
\end{equation}
  where $\chi_i$ and $\chi_f$ are spinors for the initial and final nucleon,
  and  $\vec k$ , $\vec k'_1$ and $\vec k'_2$ are the momenta of the pions
  depicted in Fig.\ref{fig:twopi}.
  Among these amplitudes,
  the upper two amplitudes which have $I_{\pi\pi}=1$  correspond to
  transitions from $\pi N$ in an S-wave,
  while the lower two amplitudes correspond to the transition
  in P-wave.

  These amplitudes have been extracted by analyzing 
  the experimental cross sections.
  In ref. \cite{manley}, 
  the data from threshold to 1470 MeV are analyzed.
  The reduced amplitudes $a_{ij}$ are assumed to be constant except for 
  $a_{10}$ which has a resonance behavior due to 
  a coupling to the Roper resonance, $P_{11}$(1440), 
\begin{equation}
   a_{10}(\sqrt s) = a'_{10} 
     \frac{M-\sqrt s_{th}}
          {M-\sqrt s - \frac{i \Gamma}{2}
         \left(\frac{\sqrt s - \sqrt s_{th}}{M - \sqrt s_{th}} \right)^2 }
\end{equation}
  where $\sqrt s_{th} \sim 1213$ MeV is the threshold energy.
  The reduced amplitudes obtained are 
\begin{equation}
\begin{array}{crccr}
  a_{11}  = &  10.61 \pm 0.62 \  [ m_{\pi}^{-3} ] &~,~&
  a_{31}  = & -6.02  \pm 0.31 \  [ m_{\pi}^{-3} ] 
\\
  a'_{10} = &  6.63  \pm 0.21 \  [ m_{\pi}^{-3} ] &~,~&
  a_{32 } = &  2.75  \pm 0.13 \  [ m_{\pi}^{-3} ] 
\end{array}
\end{equation}
  with  $M=1416 \pm 14$ MeV and $\Gamma=287 \pm 43$ MeV.
  In ref. \cite{burkhardt}, the data close to threshold, 
  including newer data with respect to \cite{manley}, are analyzed. 
  In this case the reduced amplitudes are assumed to depend linearly
  on the center of mass energy.
  The reduced amplitudes 
\begin{eqnarray}
  a_{11} &=&  3.3 \pm 0.8 
              + (0.9 \pm 2.0) (\sqrt s - \sqrt s_{th})/ m_{\pi}
  \ \ [ m_{\pi}^{-3} ] 
  \\
  a_{31} &=& -5.0 \pm 2.2 
              + (15.0 \pm 4.2) (\sqrt s - \sqrt s_{th})/m_{\pi}
  \ \ [ m_{\pi}^{-3} ] 
  \\
  a_{10} &=&  6.55 \pm 0.16 
              + (10.4 \pm 0.8) (\sqrt s - \sqrt s_{th})/m_{\pi}
  \ \ [ m_{\pi}^{-3} ] 
  \\
  a_{32} &=&  2.07 \pm 0.10 
              + (1.98 \pm 0.33) (\sqrt s - \sqrt s_{th})/m_{\pi}
  \ \ [ m_{\pi}^{-3} ] 
\end{eqnarray}
 are obtained.

\begin{figure}[t]
\begin{center}
  \includegraphics[height=70mm]{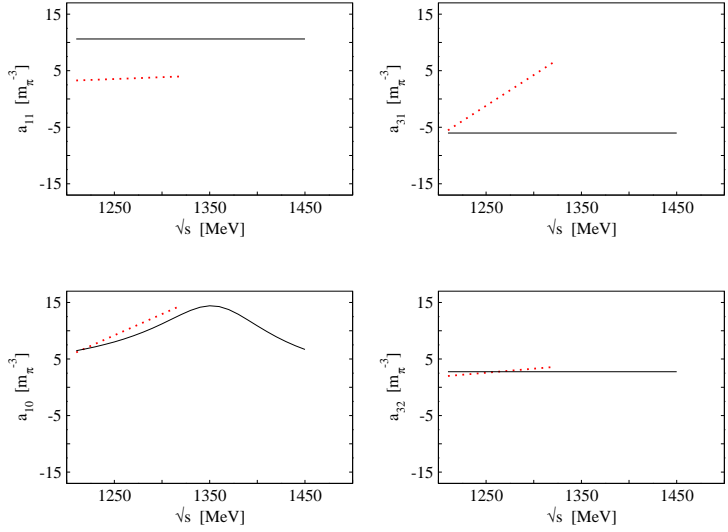}
  \caption{ 
   Empirical amplitudes of $\pi N  \to \pi \pi N$ transition.
   The continuous lines and dotted lines correspond to the
   amplitudes of paper \cite{manley} and \cite{burkhardt} 
   respectively. 
   The absolute value is plotted for $a_{10}(\rts)$ of \cite{manley}.
         }
  \label{fig:manburamp}
\end{center}
\end{figure}

 We plot both empirical reduced amplitudes in Fig.\ref{fig:manburamp}.
 One can see that the extracted amplitudes of S-wave $\pi N$
 are very different in both analyses, 
 while those of P-wave $\pi N$ agree quite well.
 These discrepancies reflect the difficulties to determine the amplitudes of 
 the S-wave $\pi N$ from $(\pi, 2\pi)$ cross section data, 
 which looks quite natural since most of the processes are dominated 
 by the transition from P-wave.
 Therefore, in this paper, we leave free
 the S-wave $\pi N$ amplitudes, $a_{11}(\rts)$ and $a_{31}(\rts)$,
 and determine them through the present study
 so that they are compatible with the data of
 $\pi N \to \pi N$ elastic scattering.

\subsection{The model including the $\pi \pi N$ channel}

 We consider the Bethe-Salpeter equation for the scattering matrix, 
 eq. (\ref{eqn:bseq}) 
 with the eight coupled channels including $\pi \pi N$, namely,
 $\pi^- p$ ,  $\pi^0 n$ , $\eta n$  ,
 $K^+ \Sigma^-$ , $K^0 \Sigma^0$ , $K^0 \Lambda$,
 $\pi^0 \pi^- p$ and  $\pi^+ \pi^- n$.
 We do not include the $\pi^0 \pi^0 n$ channel
 because it does not couple to the S-wave $\pi N$ state.

\newcommand{\sfac}
        {\chi_f^{\dagger} \vec{\sigma} \! \cdot \! (\vec k'_1 \! - \! \vec k'_2) \chi_i}
\newcommand{\pfac}{\chi_f^{\dagger} \vec{\sigma} \! \cdot \! \vec k \chi_i}

 The potentials of the $\pi N  \leftrightarrow \pi \pi N$ transitions 
 are written as 
\begin{eqnarray}
 V_{\pi^- p \, , \, \pi^0 \pi^-  p}
 &=&
    \left[ \frac{\sqrt2}{3} v_{11} + \frac{\sqrt2}{6} v_{31}\right] \sfac
  - \frac{\sqrt{10}}{10} v_{32} \pfac
 \\
 V_{\pi^- p \, , \, \pi^+ \pi^-  n}
 &=&
    \left[ \frac13  v_{11} - \frac13  v_{31} \right] \sfac  
  + \left[ \frac{\sqrt2}{3}  v_{10}- \frac{\sqrt5}{15} v_{32}  \right] \pfac 
 \\
 V_{\pi^0 n \, , \, \pi^0 \pi^-  p}
 &=&
    \left[  - \frac13  v_{11} + \frac13  v_{31} \right]  \sfac
  - \frac{\sqrt{5}}{5} v_{32} \pfac
 \\
 V_{\pi^0 n \, , \, \pi^+ \pi^-  n}
 &=&
  \left[ - \frac{\sqrt2}{6} v_{11} - \frac{\sqrt{2}}{3} v_{31} \right]  
  \sfac
  +
  \left[ - \frac13 v_{10} - \frac{\sqrt{10}}{15} v_{32} \right]  
  \pfac
\end{eqnarray}
 in terms of the reduced potentials 
 $v_{11}(\rts)$  ,  $v_{31}(\rts)$, $v_{10}(\rts)$ and $v_{32}(\rts)$,
 which correspond to the reduced amplitudes 
 $a_{11}(\rts)$  ,  $a_{31}(\rts)$  $a_{10}(\rts)$ and $a_{32}(\rts)$
 respectively, after solving the BS equation with all the channels.
 Note that, in our formalism, $\pi^+$ corresponds to the
 $-\ketv I=1,I_z=1>$ state.
 The terms of $v_{10}(\rts)$ and $v_{32}(\rts)$ do not contribute 
 in the present S-wave scattering.

\begin{figure}[t]
\parbox{.58\textwidth}
 {
  \includegraphics[height=27mm]{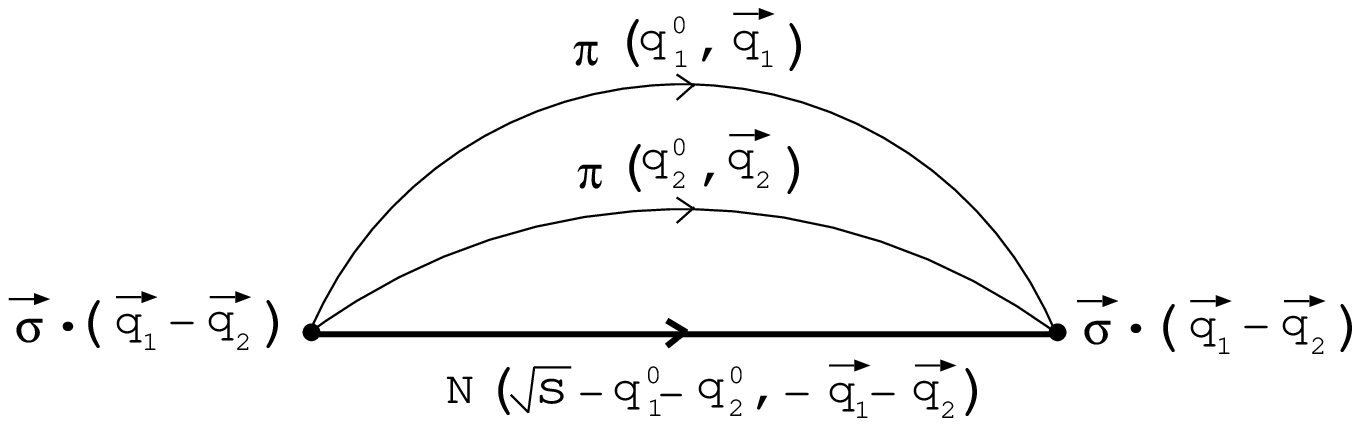}
  \caption{ Propagator for the $\pi \pi N$ state.}
  \label{fig:pipinprog}
 }
\parbox{.4\textwidth}
 {
 \includegraphics[height=40mm]{fig12} 
 \caption{ Imaginary part of $\tilde G(\rts)$.}
 \label{fig:imgtilde}
 }
\end{figure}

 For the $\pi \pi N$ state, we introduce the baryon-two meson 
 propagator corresponding to Fig.\ref{fig:pipinprog}
\begin{equation}
   \tilde G(P) = i^2 
     \int \! \! \frac{d^4 q_1}{(2 \pi)^4} 
     \int \! \! \frac{d^4 q_2}{(2 \pi)^4} 
      (\vec q_1 - \vec q_2)^2  
      \frac{2 M_N}{(P- q_1- q_2)^2 - M_N^2 + i \epsilon}
      \frac{1}{q_1^2 - m_{\pi}^2 + i \epsilon} 
      \frac{1}{q_2^2 - m_{\pi}^2 + i \epsilon} 
%
\label{eqn:gtilde}
\end{equation}
 which includes the vertex structure for convenience. 
 The integral in eq. (\ref{eqn:gtilde}) is strongly divergent.
 Its Imaginary part is finite and,
 neglecting the contribution of the negative energy of the baryon, 
 is given by, 
\begin{equation}
 \mbox{Im}[\tilde G(\rts)] = 
   - \frac{M_{N}}{4 (2 \pi)^3}
   \int \! \! d \omega_1  \int \! \! d \omega_2
   \left[ M_N^2 + 2 q_1(\omega_1)^2 + 2 q_2(\omega_2)^2
        - (\rts - \omega_1  - \omega_2)^2 \right]
                 \theta(1 - A^2)
\end{equation}
where
\begin{equation}
 A = \frac{ (\rts - \omega_1 - \omega_2)^2 - M_N^2 
               - q_1(\omega_1)^2 - q_2(\omega_2)^2 }
          { 2 q_1(\omega_1) q_2(\omega_2)}
\end{equation}
 with $q_i(\omega_i)=\sqrt{ \omega_i^2 - m_{\pi}^2}$. 
 This imaginary part, 
 reflecting the phase space for the intermediate baryon-two meson system, 
 is shown in Fig.\ref{fig:imgtilde}.
 On the other hand, 
 the real part of $\tilde G(\rts)$, in the renormalized model, is not fixed.
 The situation is well illustrated in a dispersion relation 
\begin{eqnarray}
  \mbox{Re}[\tilde G(\sqrt s)] 
  &=& \frac{(\sqrt s- w_0)^6}{\pi}  \int_{\sqrt s}^{w_{max}} \! \! d w
      \frac{\mbox{Im}[\tilde G(w)]}{(w-w_0)^6(w-\sqrt s)}
      + A + B(\sqrt s-w_0) 
  \nonumber
  \\ 
  & &
     + C(\sqrt s-w_0)^2
     + D(\sqrt s-w_0)^3 + E(\sqrt s-w_0)^4 + F(\sqrt s-w_0)^5
\end{eqnarray}
 where six arbitrary constants $A \sim F$ are introduced
 because Im$[\tilde G(\rts)]$ grows as 
\begin{equation}
 \lim_{\sqrt s \to \infty} 
  \frac{\mbox{Im}[\tilde G(\sqrt s)]}{\sqrt s^n} \to 
 \left\{
 \begin{array}{ccc}
     - \infty  & \quad \mbox{for} & \quad n \le 5 \\
        0    & \quad \mbox{for} & \quad n \ge 6
 \end{array}
 \right.
\end{equation}
 which is easily proved by an actual calculation
 or dimensional considerations. Therefore, in this paper, 
 we treat $\mbox{Re}[\tilde G(\rts)]$ as a free function
 and look for reasonable values consistent with the experiment.

\subsection{Final results}

 By varying the potential $v_{11}(\rts)$ and $v_{31}(\rts)$, 
 the subtraction constants $a_i(\mu)$ and Re$[\tilde G(\rts)]$,
 we try to reproduce the experimental elastic $\pi N$ T matrix
 and also the $\pi N \to \pi \pi N$ cross sections.
 For the $v_{11}(\rts)$ and $v_{31}(\rts)$ functions 
 we employ a real polynomial function, 
 and for Re[$\tilde G(\rts)$] we test several types of functions.

\begin{figure}[t]
\begin{center}
  \includegraphics[height=120mm]{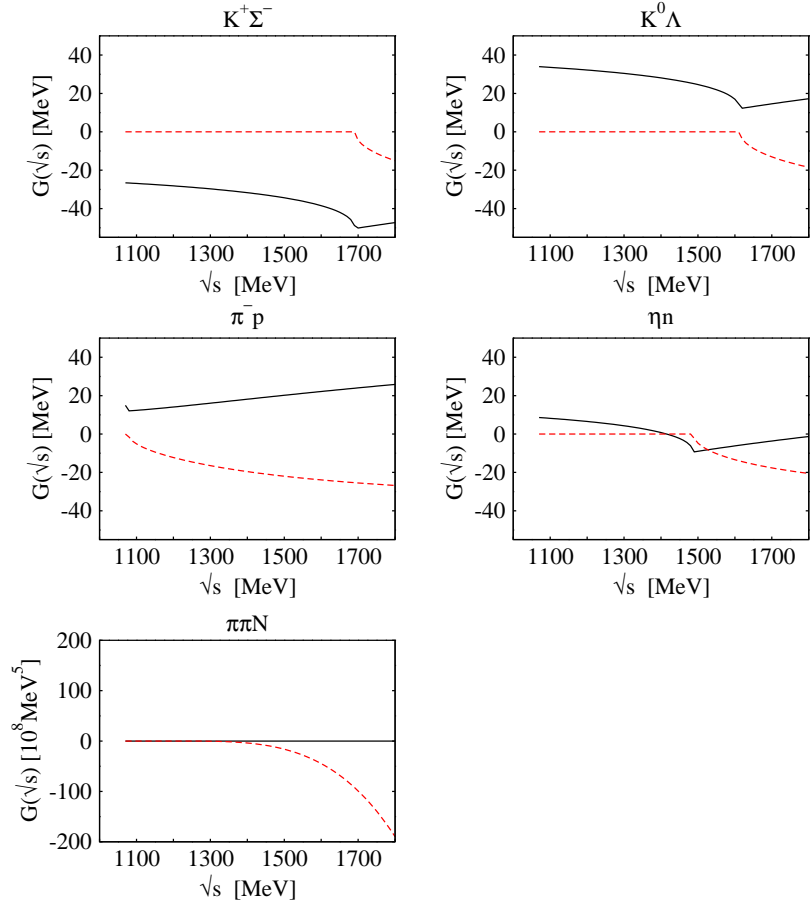}
  \caption{ 
  Propagators of meson-baryon and $\pi\pi N$ systems
  using the parameters of eq (\ref{eqn:subpipin}).
  The solid and dashed lines are real and imaginary parts respectively.
          }
  \label{fig:propall}
\end{center}
\end{figure}

 In Fig.\ref{fig:propall} we show the propagators 
 $G(\rts)$ and $\tilde G(\rts)$. 
 The subtraction parameters $a_i(\mu)$ for the meson-baryon propagators
 that we obtain in the new fit to the data, are
\begin{equation}
   \mu= 1200 ~\mbox{MeV},~~ 
   a_{\pi N}(\mu)     =  2.0 ,~~
   a_{\eta N}(\mu)    =  0.1 ,~~
   a_{K \Lambda}(\mu) =  1.5 , ~~ 
   a_{K \Sigma}(\mu)  = -2.8
\label{eqn:subpipin}
\end{equation} 
 which are a little changed from the previous values
 omitting the $\pi \pi N$ channels.
 For the $\tilde G(\rts)$ case the real part is taken zero.
 We find that the best results are obtained
 with functions compatible with zero.
 This result is similar to the one found in \cite{klingl}
 where the real part of the three pion loop form $\phi$ decay 
 was also found negligibly small.

\begin{figure}[t]
\begin{center}
  \includegraphics[height=80mm]{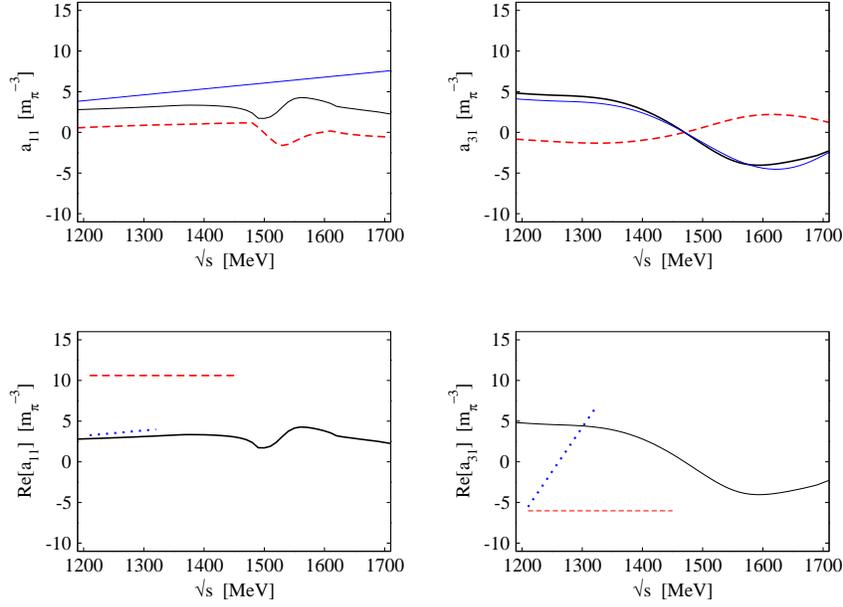}
  \caption{ 
 The S-wave $\pi N\leftrightarrow \pi \pi N$ amplitudes 
 $a_{11}(\rts)$ and $a_{31}(\rts)$. 
 Upper figures in the panel: the thin lines are the potential $v_{ij}(\rts)$,
 the solid lines are Re$[a_{ij}(\rts)]$ and the dashed lines are Im$[a_{ij}]$.
 Lower figures in the panel: the calculated Re$[a_{ij}]$(solid) are compared
 with \cite{manley}(dashed) and \cite{burkhardt}(dotted).
         }
  \label{fig:a11a31ino}
\end{center}
\end{figure}

 Fig.\ref{fig:a11a31ino} shows the functions $a_{11}(\rts)$ and 
 $a_{31}(\rts)$ determined in this study. In the upper two graphs, 
 the potentials $v_{11}(\rts)$ ($v_{31}(\rts)$) and 
 the amplitudes $a_{11}(\rts)$ ($a_{31}(\rts)$), 
 which come after unitarization, are shown.
 The thin lines correspond to the following potentials which we finally use
\begin{eqnarray}
  v_{11}(\rts)&=&4.0 + 1.0 (\sqrt s - 1213)/ m_{\pi}  ~~[ m_{\pi}^{-3} ] 
  \\
  v_{31}(\rts)&=&-5.60 (\sqrt s - 1470)/m_{\pi} 
                 -1.05 (\sqrt s - 1470)^2/m_{\pi}^2 
                 +1.77 (\sqrt s - 1470)^3/m_{\pi}^3 
              \\
              & &
                 +0.66 (\sqrt s - 1470)^4/m_{\pi}^4 
                 -0.17 (\sqrt s - 1470)^5/m_{\pi}^5 
                 -0.07 (\sqrt s - 1470)^6/m_{\pi}^6   ~~[ m_{\pi}^{-3} ] ~~.
              \nonumber
\end{eqnarray}
 The continuous line and dashed line correspond to the real and imaginary part 
 of $a_{11}(\rts)$ ($a_{31}(\rts)$) respectively.
 One can see that the imaginary parts are almost negligible
 as assumed in the paper \cite{manley} and \cite{burkhardt}.
 In the lower two graphs, the real part of $a_{11}(\rts)$ ($a_{31}(\rts)$) 
 is plotted in comparison with the two empirical ones.
 The lower energy part of the calculated $a_{11}(\rts)$ agrees with that of 
 the paper \cite{burkhardt} but it is quite different from the one of the
 \cite{manley}. On the other hand our calculated $a_{31}(\rts)$ amplitude
 is different from both \cite{manley} and \cite{burkhardt}. 
 We should note that,
 while it is possible to reproduce the $\pi N \to \pi \pi N$ cross sections
 with the three set of amplitudes, a simultaneous description of the 
 $\pi N \to \pi \pi N$ cross sections and the $\pi N \to \pi N$ scattering
 date is not possible with the amplitude of \cite{manley} and \cite{burkhardt}.
 We shall elaborate further on this issue below.

\begin{figure}[p]
\begin{center}
  \includegraphics[height=80mm]{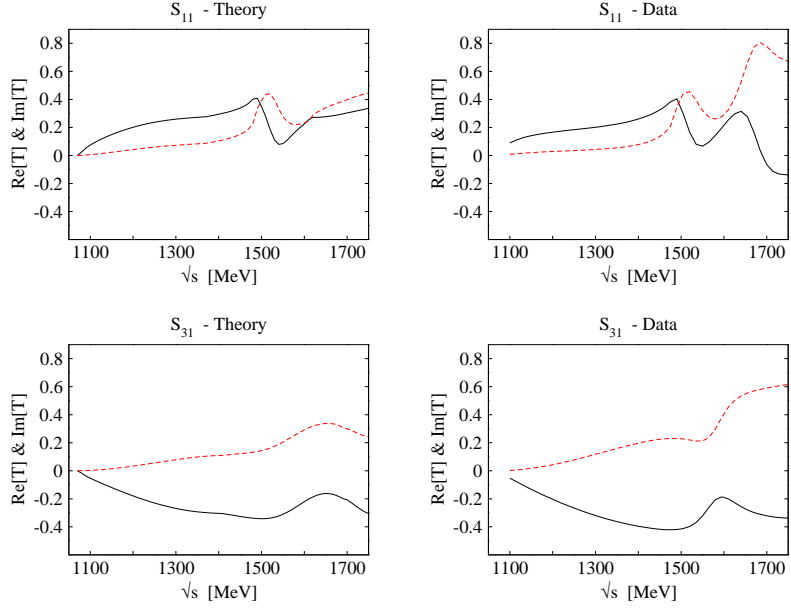}
  \caption{
    Scattering amplitude for the $S_{11}$ and $S_{31}$ $\pi N$ 
    partial waves with $\pi\pi N$ channels.
          }
  \label{fig:tmatfull}
\end{center}
\end{figure}

\begin{figure}[p]
\begin{center}
  \includegraphics[height=100mm]{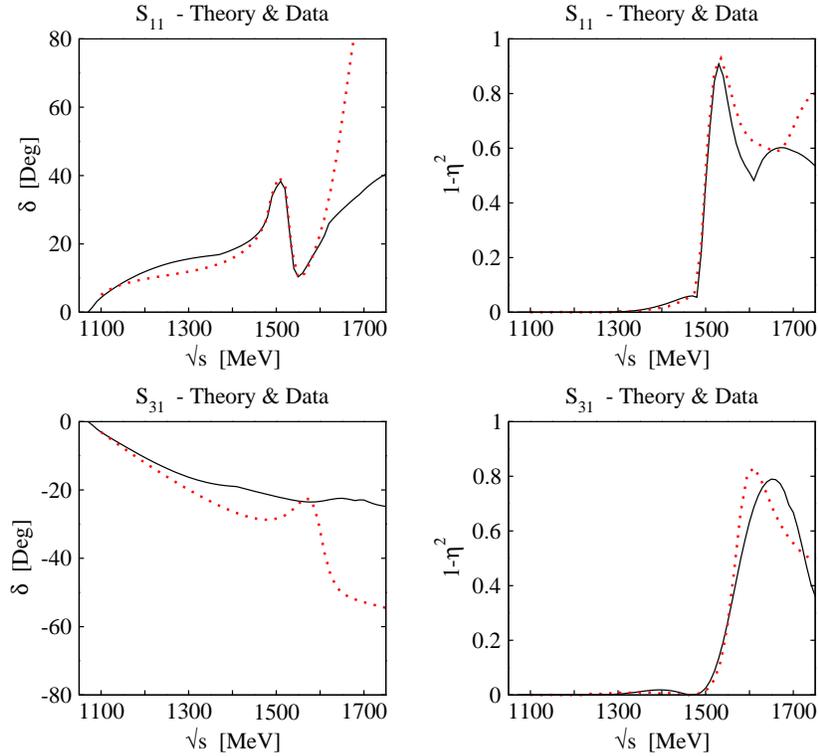}
  \caption{ 
    Phase-shifts and inelasticities of 
    $S_{11}$ and $S_{31}$ $\pi N$ scattering with $\pi \pi N$ channels.
          }
  \label{fig:phinfull}
\end{center}
\end{figure}

\begin{figure}[t]
\begin{center}
  \includegraphics[height=50mm]{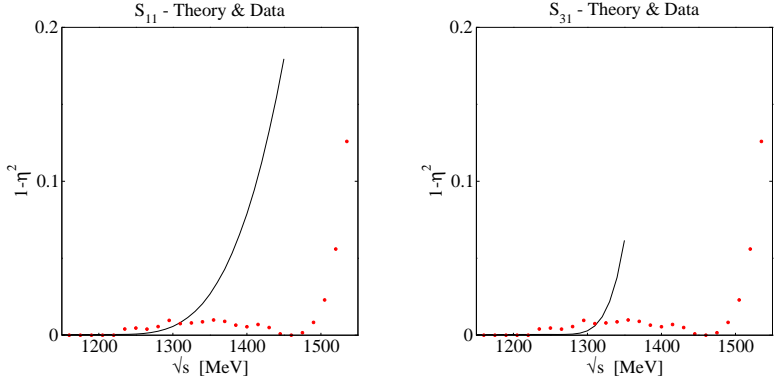}
  \caption{
     Inelasticities of $S_{31}$ $\pi N$ scattering which 
     correspond to the empirical $\pi N \to \pi \pi N$ amplitudes.
     The solids line in the left and right figure 
     correspond to \cite{manley} and \cite{burkhardt} respectively.
     The dotted line is the experimental analysis \cite{CNS}.
          }
  \label{fig:s31inemp}
\end{center}
\end{figure}

 The T matrix elements obtained are shown in Fig.\ref{fig:tmatfull}.
 One can see that the T matrix elements are well reproduced 
 at energies below 1600 MeV in both the $S_{11}$ and $S_{31}$ cases.
 The phase-shifts and inelasticities are shown in Fig.\ref{fig:phinfull}.
 The inclusion of the $\pi \pi N$ channels improves slightly the
 phase shifts in $S_{11}$ and $S_{31}$.
 The important thing to note is that, as seen in Fig.\ref{fig:phinfull},
 the inelasticities at low energies in both $S_{11}$ and $S_{31}$ are well 
 reproduced with the inclusion of the $\pi \pi N$ channels.
 The present $a_{11}(\rts)$ function is essentially the same as that of 
 the paper \cite{burkhardt} in the range of energies studied there.
 Should we take the amplitude of paper \cite{manley},
 the inelasticities would be much overestimated at energies around 1400 MeV.
 The present $a_{31}(\rts)$ amplitude is different from both empirical
 $a_{31}(\rts)$ determinations.
 It has a node at the energy 1470 MeV which is reflected 
 in the inelasticities shown  in the figure.
 The inelasticities are quite sensitive to the $a_{31}(\rts)$ amplitude used.
 For example, the two empirical $a_{31}(\rts)$ correspond to the inelasticities 
 shown in Fig.\ref{fig:s31inemp} which differ appreciable from those obtained
 with the function determined here.
 We should note that for this test we have used a function $v_{31}(\rts)$ such
 that after unitarization they lead to scattering amplitudes identical to the 
 empirical $a_{31}(\rts)$.
 The unitarization procedure modifies only little the potential 
 as is visible in upper right box of Fig.\ref{fig:a11a31ino} for our case.
 The opposite sign of our $a_{31}(\rts)$ also
 provides the same inelasticities and hence 
 is compatible with the $S_{31}$ $\pi N$ scattering data.
 However, it leads to unacceptable results for the  
 $\pi N \to \pi \pi N$ cross section.

\begin{figure}[p]
\begin{center}
  \includegraphics[height=120mm]{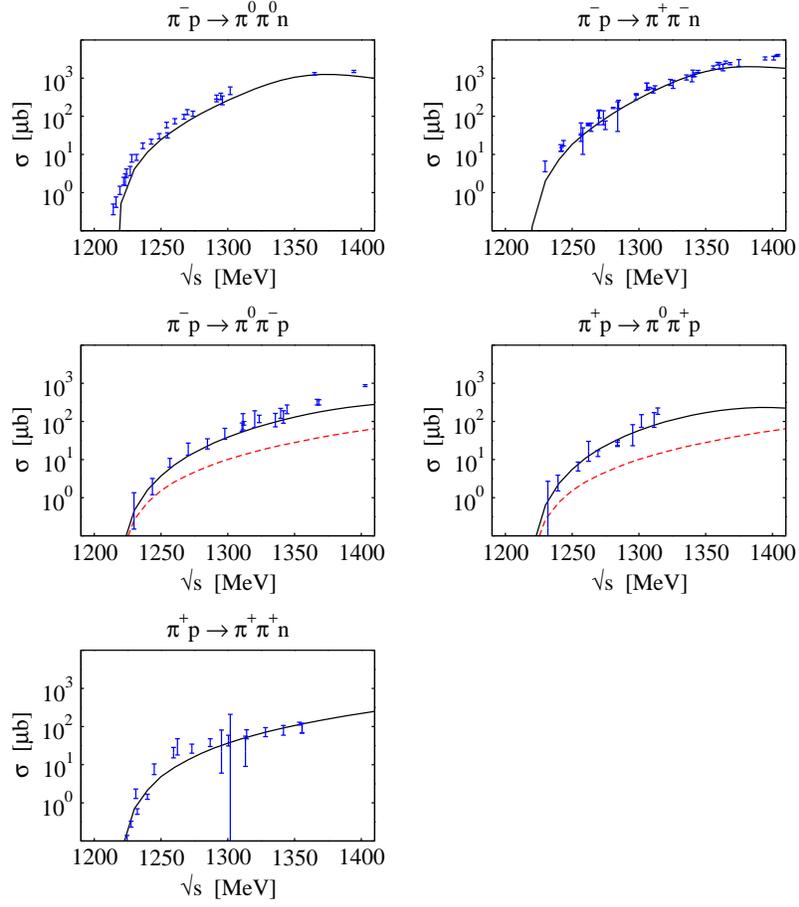} 
  \caption{ The cross section of $\pi N \to \pi \pi N$ scattering.
            Data points are taken from the papers
            in reference of \cite{burkhardt}.}
  \label{fig:twocrfull}
\end{center}
\end{figure}

\begin{figure}[p]
\begin{center}
  \includegraphics[height=40mm]{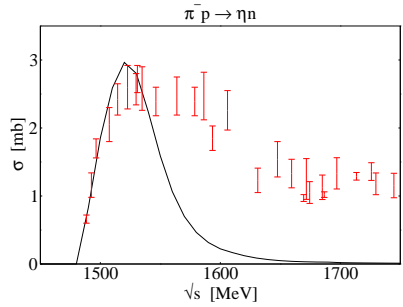} 
  \caption{ The cross section of $\pi N \to \eta n$ scattering.}
  \label{fig:etan}
\end{center}
\end{figure}

 Fig.\ref{fig:twocrfull} shows the $\pi N \to \pi \pi N$ cross sections.
 They are calculated with the present S-wave amplitudes 
 and the P-wave amplitudes of the paper \cite{manley}. 
 The dashed lines correspond to the cross section when we drop the S-wave 
 contributions. The two processes, 
 $\pi^- p \to \pi^0 \pi^0 n$ and $\pi^+ p \to \pi^+ \pi^+ n$,
 are purely P-wave $\pi N$ and have no S-wave $\pi N$ contribution.
 In the $\pi^- p \to \pi^+ \pi^- n$ reaction,
 the P-wave contribution is large and dominates the process.
 The effect of the S-wave is not visible in the figure. However, 
 in the $\pi^- p \to \pi^0 \pi^- p$ and $\pi^+ p \to \pi^0 \pi^+ p$ processes
 we can see that the P-wave contributions are small and do not explain the 
 size of the data.
 As one can see, the present S-wave amplitude provides enough strength 
 to account for the cross section data.
 In the $\pi^- p \to \pi^0 \pi^- p$ process, 
 the S-wave amplitude is a mixture of $a_{11}(\rts)$ and $a_{31}(\rts)$.
 In order to obtain the contribution of the S-wave shown in the figure, 
 the amplitudes $a_{11}(\rts)$ and $a_{31}(\rts)$ should
 interfere constructively close to the $\pi \pi N$ threshold.
 The relative sign between $a_{11}(\rts)$ and $a_{31}(\rts)$
 ($v_{11}(\rts)$ and $v_{31}(\rts)$) is determined by this condition.
 The value of $a_{11}/a_{31}$ at threshold is about $+1/2$, 
 which is in contrast to the one pion exchange prediction, $-2$ \cite{Aaron}.
 However, more elaborate models for two pion production
 \cite{manolo,vesna,dillig,bernard1,bernard2,miranda} contain many more terms
 that change the value of this ratio. 
 The amplitudes $a_{11}(\rts)$ and $a_{31}(\rts)$ determined in this study,
 account for both the $\pi N$ elastic scattering data and 
 the $\pi N \to \pi \pi N$ cross section data simultaneously.

 In Fig.\ref{fig:etan} we also show the result for the
 $\pi^- p \to \eta n$ cross section.
 We can see that the agreement is good up to 1550 MeV,
 above which higher partial waves, as demonstrated in \cite{caro},
 become relevant.  

 We have also evaluated the scattering lengths in our approach which are 
 listed in Table \ref{tbl:length}. The thresholds for the
 \chthr, \chtwo, \chone are $\rts=$ 1613, 1690, 1691 MeV
 respectively.  These energies, particularly the two last ones,
 are already in the region of energies where the theory deviates
 from experiment in the phase shifts and inelasticities,
 therefore we should take these numbers only as indicative.
 On the other hand, the fitting to the data has been done around the 
 $N^*$(1535) energy. Thus, our prediction for the $\eta n$
 scattering length, $a_{\eta n}=0.26 + i 0.25$ fm, should be rather accurate.
 This number is in agreement with the result quoted in \cite{kaiser},
 $a_{\eta n}=0.20 + i 0.26$ fm, although a bit more attractive.
 Still the real part is about a factor three smaller than in \cite{wycech} 
 and a factor four smaller than in \cite{svarc}.
 In spite of that, it was argued in \cite{kaiser} that
 even the small value $\mbox{Re}[a_{\eta n}] = 0.2$ fm is not
 unrealistically small.
 This scattering length is important since it plays a crucial role in the 
 possibility to have $\eta$ bound states in nuclei \cite{chiang,zaki}.
 The scattering length for $\pi^0 n$  and $\pi^- p$ have not been imposed
 in the fit to the data, 
 which concentrated in the $N^*$(1535) region, as already mentioned.
 In this sense, the agreement with the data about 450 MeV below that resonance
 should be considered an unexpected success. 
 We obtain isospin 3/2 and isospin 1/2 scattering lengths
 $a_3=-0.0875 \,m_{\pi}^{-1}$, $a_1=0.1272 \,m_{\pi}^{-1}$,
 to be compared with the experimental numbers \cite{schroder},
 $a_3^{exp}=-0.0852 \pm 0.0027 \,m_{\pi}^{-1}$, 
 $a_1^{exp}= 0.1752 \pm 0.0041 \,m_{\pi}^{-1}$.
 The agreement is good for the isospin 3/2 scattering length,
 but the isospin 1/2 one is about 25\% smaller than experiment.

\begin{table}[t]
\begin{center}
\caption{ The calculated meson-baryon scattering lengths in unit of [fm]. } 
\label{tbl:length}
\begin{tabular}{cccc}
\hline
            & \chfiv
            & \chfou
            & \chsix 
 \\
 \hline     
               $a_i$ [fm]
            & -0.023          
            &  0.080 + i 0.003
            &  0.264 + i 0.245 
 \\
 \hline
 \hline     

            & \chthr          
            & \chtwo 
            & \chone
 \\
 \hline     
               $a_i$ [fm]
            & -0.148 + i 0.165 
            & -0.205 + i 0.068 
            & -0.284 + i 0.090
 \\
 \hline
\end{tabular}
\end{center}
\end{table}

 To summarize the result of section 2 and 3, wee can see that
 the chiral unitary approach including the $\pi \pi N$ channels,
 is a very efficient tool to study the S-wave $\pi N$ scattering. 
 It can reproduce, with few parameters, 
 not only the isospin 1/2 part but also the isospin 3/2 part,
 which could not be obtained before \cite{siegel,hosaka,juan},
 when only meson-baryon channels were considered.
 Particularly, the $\pi \pi N$ channel was found
 essential to reproduce the isospin 3/2 scattering.

\section{The $N^*$(1535) and $\Delta$(1620) resonances}

\begin{figure}[t]
\begin{center}
  \includegraphics[height=60mm]{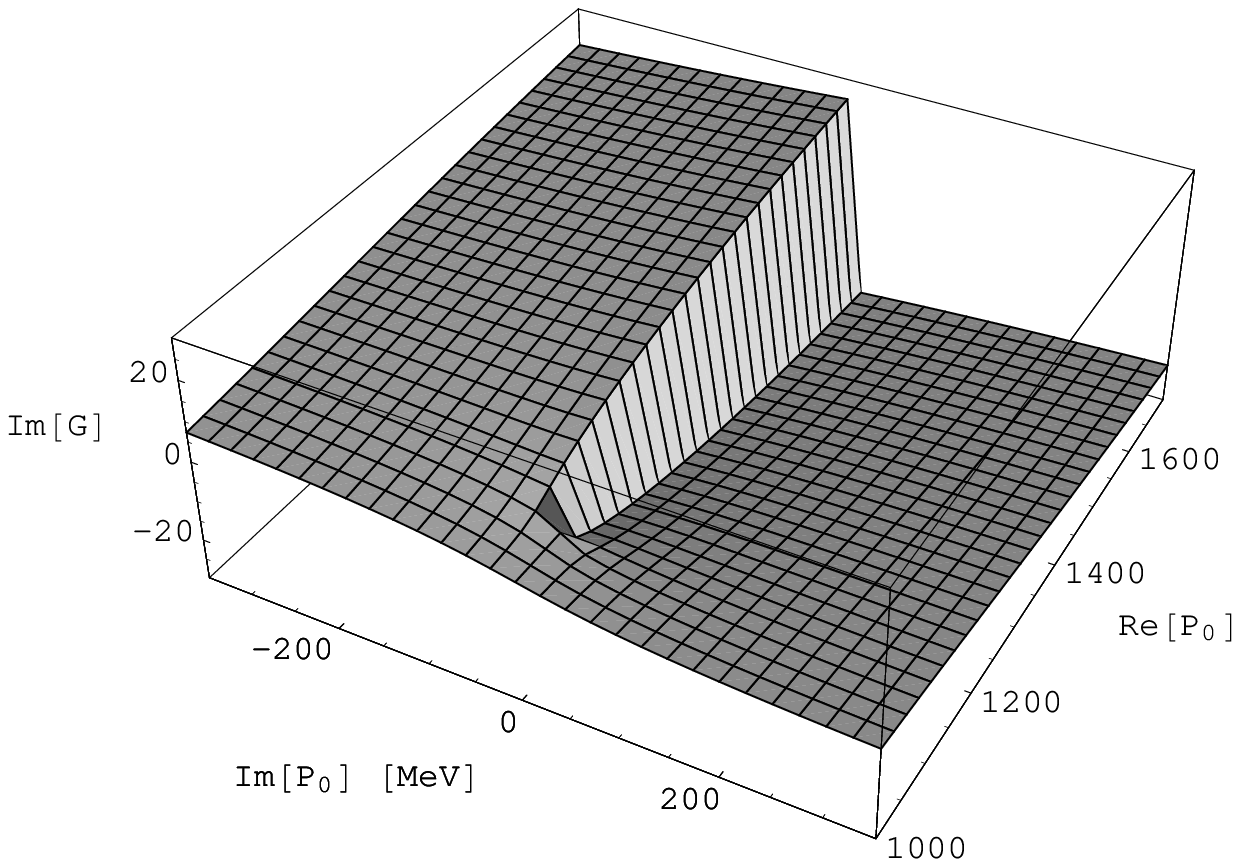} 
  \includegraphics[height=60mm]{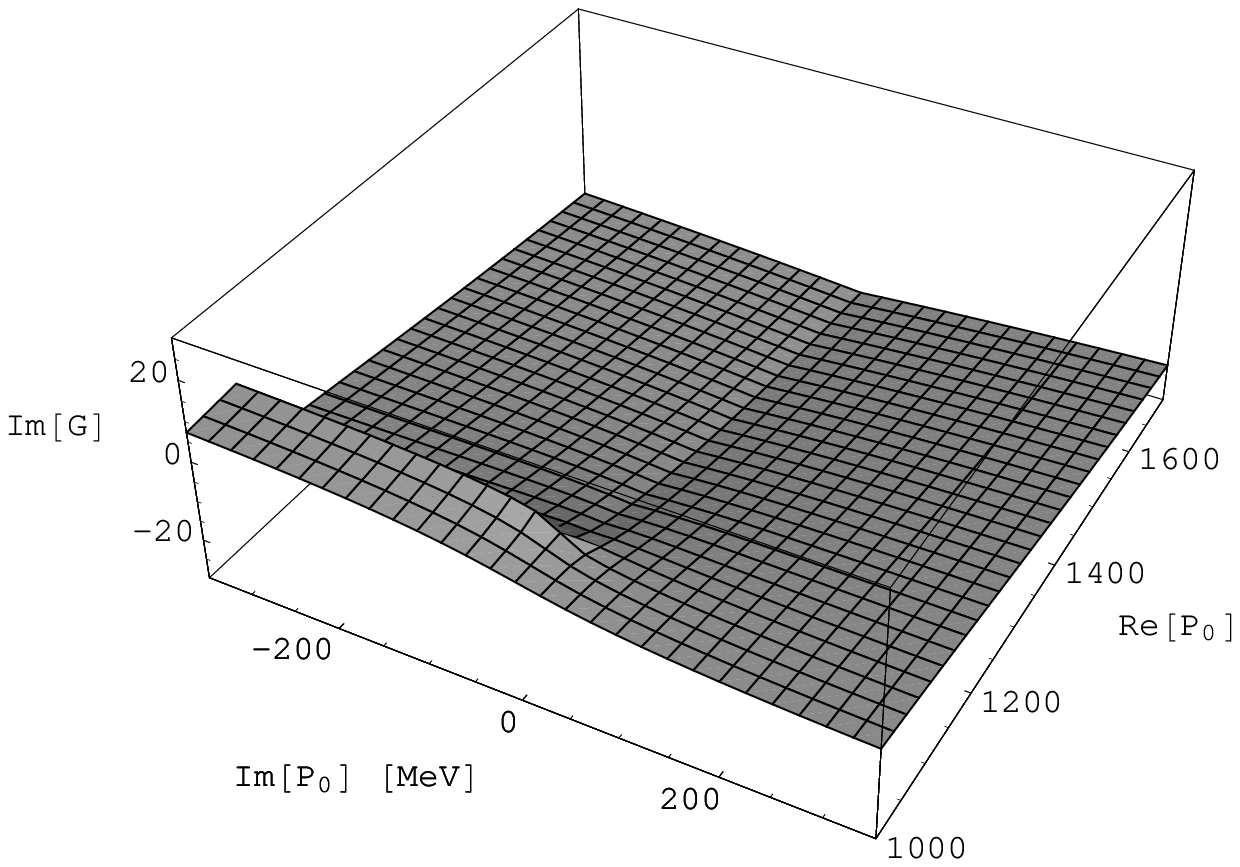} 
  \caption{The imaginary part of the propagator for the $\pi^- p$ system. }
  \label{fig:cmprop}
\end{center}
\end{figure}

 As shown in Fig.\ref{fig:tmatfull}, 
 the calculated $T(S_{11})$ has a resonance behavior around 1535 MeV
 indicating that the well known negative parity baryon $N^*(1535)$ 
 is generated in this approach.
 To find the pole corresponding to the resonance
 we extend our calculation of the T matrix to the complex $P_0$ plane.
 We evaluate the T matrix elements by means of
\begin{equation}
  T(P_0) = \left[ V(P_0)^{-1} - G(P_0) \right]^{-1}
\end{equation}
 and look for poles in the complex $P^0$ plane. 
 In this plane the function $G(P_0)$ has cuts on the real axis.
 For example, Fig.\ref{fig:cmprop} left shows
 $\mbox{Im}[G_{\pi^- p}(P_0)]$ in the physical sheet,
 namely the first Riemann sheet.
 We also plot to the right of Fig.\ref{fig:cmprop} the sheet
 connecting the first Riemann sheet for Im$[P_0]\ge 0$ 
 with the second Riemann sheet for Im$[P_0]<0$, which is defined as
\begin{equation}
  G_i^{II}(P_0) =  
 \left\{
 \begin{array}{lcc}
  G_i^{I}(P_0)   & 
  ~ \mbox{at} & \mbox{Re}[P_0] \le \sqrt{s^0}_i
  \\
  G_i^{I}(P_0) - 2 i \mbox{Im}[G_i^{I}(P_0)] & 
  ~ \mbox{at} & \mbox{Re}[P_0]  >  \sqrt{s^0}_i
 \end{array}
 \right.
\end{equation}
 with $\sqrt{s^0}_i$ the $i$-th channel threshold energy .
 We need also to extrapolate $\tilde G(\rts)$ to the complex plane.
 For this purpose we parameterize the result for Im$[\tilde G(\rts)]$
 above the $\pi \pi N$ threshold in the real axis as
\begin{equation}
 \mbox{Im}[\tilde G(\rts)] =
        -0.638(\rts - 1213)/m_{\pi}
        +1.124(\rts - 1213)^2/m_{\pi}^2
        -0.882 (\rts - 1213)^3/m_{\pi}^3
\end{equation}
 which allows for an analytical continuation 
 for Re$[P_0]$ above that threshold.

\begin{figure}[t]
\begin{center}
  \includegraphics[height=60mm]{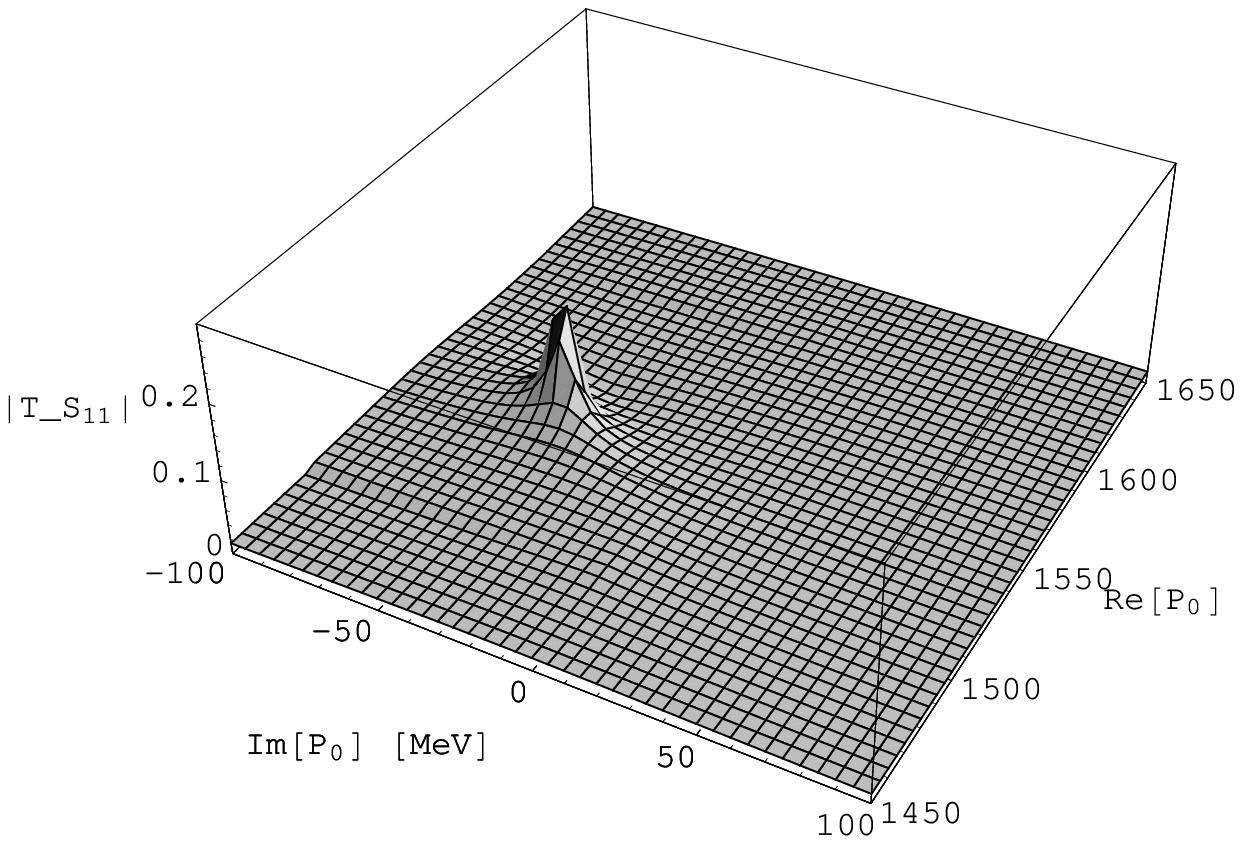} 
  \includegraphics[height=60mm]{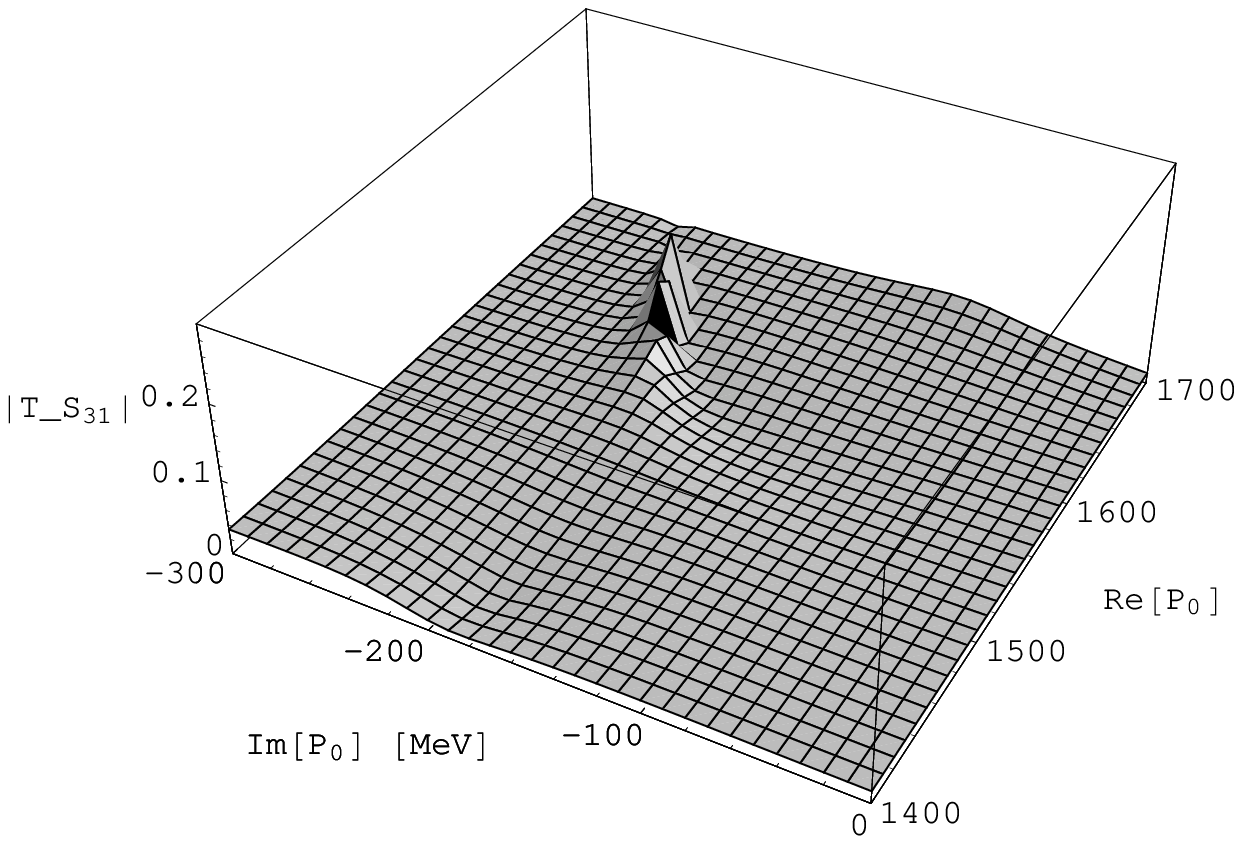} 
  \caption{ 
    $|T(S_{11})|$ (left) and $|T(S_{31})|$ (right) 
    in the complex energy plane.
    The poles of $N^*(1535)$ and $\Delta(1620)$ are seen
    at the energies  $1543 - i 46$ MeV and $1625 - i 215$ MeV respectively.
        }
  \label{fig:pole}
\end{center}
\end{figure}

 We search for poles of the isospin 1/2 T matrix elements on this sheet,
 and obtain a pole at
\begin{equation}
 P_0^R \equiv 1543 - i~ 46 ~\mbox{MeV} ~~.
\end{equation}
 The structure of the T matrix guarantees that we find
 the pole in all the elements which have an isospin 1/2 component.
 For example, Fig.\ref{fig:pole} left shows  $|T(S_{11})|$,
 where we can see a pole clearly.
 This result tells us that the decay width of the resonance is about 93 MeV.
 This value is smaller than the PDG estimation $100 \sim 250$ MeV \cite{PDG},
 but agrees with the new data from BEPC, $95 \pm 15$ MeV \cite{Zou}.

 On the other hand, in the elements of pure isospin 3/2 we find a pole around
 $1625 - i 215$ MeV, which is reminiscent of the $\Delta$(1620) resonance, 
 although with a larger width than the nominal one of the PDG of
 about 120-160 MeV.
 In Fig.\ref{fig:pole} right we show $|T(S_{31})|$, 
 where a pole is clearly visible.
 We should consider the position of the pole only as qualitative
 in our approach since around $\rts=1600$ MeV is where
 we start having noticeable discrepancies with the data
 and furthermore we had to go far in the complex plane where
 the analytic extrapolation of the $\pi \pi N$ propagator becomes less accurate.
 This resonance is responsible for the change of curvature in Im$T(S_{31})$
 shown in Fig.\ref{fig:tmatfull}. 
 It is interesting to note that the introduction of the $\pi \pi N$ channels
 is what has induced the appearance of this resonance,
 which does not show in our approach when the $\pi \pi N$ channels
 are not considered.
 In this respect it is interesting to note that the $\Delta$(1620) resonance 
 couples mostly to the $\pi \pi N$ channel \cite{PDG}.

 For $P_0$ near the pole the T matrix elements are approximated by
\begin{equation}
  T_{ij}(P_0) \simeq \frac{g_i g_j}{P_0-P_0^R}
\end{equation}
 with $g_i$ the couplings of the resonance to the $i$-th channel. 
 We obtain the size of the couplings by evaluating the residues 
 of the diagonal elements $T_{ii}(P_0)$.
 The values obtained are listed in Table \ref{tbl:coupling}.
 We find that the resonance $N^*(1535)$ couples strongly 
 to the $K \Sigma$ and $\eta n$ channels strongly with a couplings
 four times bigger than for the $\pi N$ channels.
 The couplings to the $K \Lambda$ are also large compared
 to the $\pi N$ channels. In short, we get
\begin{equation}
 |g_{\pi N}| < |g_{K \Lambda}| < |g_{\eta n}| \sim |g_{K \Sigma}| ~.
\end{equation}

 Using these couplings 
 we calculate the partial decay widths of the open channels by means of  
\begin{equation}
 \Gamma_i =  - 2 \mbox{Im}[ G_i(M_{N^*}) ] |g_i|^2
\end{equation}
 where we take $M_{N^*} = \mbox{Re}[P_0^R] = 1543$ MeV.
 The partial decay rates and the branching ratios obtained
 are also listed in Table \ref{tbl:coupling}.
 The calculated branching ratios of $\pi N$, $\eta N$ and $\pi \pi N$ decay modes,
 are 22\%, 70\% and 7\% respectively.
 The fraction of $\eta N$ mode is large, 
 which is know to be characteristic of this resonance
 although our fraction is bigger than the PDG estimation 30 $\sim$ 55 \%.
 Our $\pi N$ fraction is smaller than the PDG estimation, 35 $\sim$ 55 \%,
 while the $\pi \pi N$ fraction is compatible with the PDG estimation, $\le$ 10 \%.

\begin{table}[t]
\begin{center}
\caption{Coupling constants and decay widths of $N^*(1535)$. } 
\label{tbl:coupling}
\begin{tabular}{ccccccccc}
 \hline
            & \chone & \chtwo & \chthr & ~\chfou~ & ~\chfiv~ & ~\chsix~ & \chsev & \cheig
 \\
 \hline
 $|g_i|$    &  2.12  &  1.50  &  0.92  &  0.56  &  0.39  &  1.84  &  0.57 $m_{\pi}^{-2}$  &  0.40 $m_{\pi}^{-2}$
 \\
 $\Gamma_i$ [MeV] &  &        &        &  14.1  &   7.0  &  65.7  &  4.6   &   2.4 
 \\ 
 B. ratio [\%]    &  &        &        &  15.0  &   7.4  &  70.1  &  4.9   &   2.5
 \\
 \hline
 \hline
 $g_i$      & -2.36  &  1.67  & -1.28  & -0.57  &  0.39  &  1.77  & -0.61 $m_{\pi}^{-2}$ & -0.43 $m_{\pi}^{-2}$
 \\ 
 $\Gamma_i$ [MeV] &  &        &        &  15.0  &   7.1  &  60.8  &  5.4   &   2.6 
 \\ 
 B. ratio [\%]    &  &        &        &  16.5  &   7.8  &  66.9  &  5.9   &   2.9
 \\
 \hline
\end{tabular}
\end{center}
\end{table}

 The lower part of Table \ref{tbl:coupling} shows the same quantities
 obtained from a Breit-Wigner fit of the real energy scattering 
 amplitudes $T_{ij}(\rts)$.
 We fit them by the Breit-Wigner form together with a background
\begin{equation}
  T_{ij}(\rts) = g_i g_j \frac{1}{\rts - 1543 + i 46} 
                + a_{ij} + b_{ij}( \rts - 1450)  
\end{equation}
 at $1450 \le \rts \le 1650$ MeV.
 The unknown parameters $g_i \times g_j$, $a_{ij}$ and $b_{ij}$ are determined 
 by the method of least squares. 
 We obtain the values of $g_i$ from the $g_{\eta n} \times g_i$ 
 corresponding to the $\eta n$ final state amplitudes. 
 Their absolute value agree fairly well with the values obtained from the pole 
 residues and gives us confidence about our numerical evaluation of the couplings.
 For instance, the $\pi N$, $\eta n$ and $\pi \pi N$ branching ratios
 are now 24 \%, 67 \% and 9 \% respectively.

 This latter analysis allows us to determine the sign of the couplings.
 The signs given are relative to that of $g_{\eta n}$.
 It is instructive to decompose the resonance in the SU(3) representations.
 In fact, our result, ignoring the coupling to the $\pi \pi N$ channel, 
 leads to
\begin{equation}
  g_{8}   = -2.52 ~,~
  g_{8'}  =  2.62 ~,~
  g_{10}  =  0.43 ~,~
  g_{27}  = -0.47 
\end{equation}
 and tells us that the $N^*(1535)$ resonance is almost an equal weight mixture
 of the R-parity even SU(3) octet $8$ and the R-parity odd SU(3) octet $8'$.
 It would be interesting to compare these results with the results from other 
 models or lattice QCD simulations.

\section{Conclusion}

 We have studied the S-wave $\pi N$ scattering, 
 together with that of coupled channels, 
 in a chiral unitary model in the region of center of mass energies from 
 threshold to 1600 MeV.
 We calculated the T matrix using the Bethe-Salpeter equation 
 in the eight coupled channels including six meson-baryon channels and 
 two $\pi \pi N$ channels.
 We took the transition potentials between the meson-baryon systems
 from the lowest order chiral Lagrangian
 and improved them taking into account the vector meson dominance hypothesis.
 Then we introduced the appropriate $\pi N \leftrightarrow \pi \pi N$
 transition potentials which influence both the elastic scattering
 and the pion production processes.
 In the present model the renormalization due to higher order contributions
 is included by means of subtraction constants in the real part of the
 propagators of the two or three-body systems,
 which are taken as free parameters and determined through comparison
 with the T matrix of the data analysis.
 The imaginary part of the meson-baryon or $\pi \pi N$ propagators
 is fixed and ensures unitarity in the S matrix.
 A realistic T matrix is obtained with a few free parameters
 for energies up to 1600 MeV.
 The phase-shifts and the inelasticities are well reproduced in both isospin 
 1/2 and 3/2.
 We find that the correction of the chiral coefficient 
 and the $\pi \pi N$ channels are important  
 to obtain an accurate T matrix, especially in isospin 3/2.
 Our analysis allowed us to determine the S-wave amplitudes for
 $\pi N \to \pi\pi N$ and we found that the isospin 3/2 
 $\pi N \leftrightarrow \pi \pi N$ amplitude is different
 from the two previous empirical ones.

 The resonance $N^*(1535)$ is generated dynamically and qualifies as a quasi 
 bound state of meson and baryon.
 The corresponding pole is seen in the T matrix on the complex plane.
 We calculate the total and partial decay width of the resonance.
 The total width obtained, about 80 MeV, is smaller than the PDG estimation,
 but agrees with the new data from BEPC.
 Also the large $\eta n$ branching ratio observed in the data is reproduced.
 
 The present study has served to show the potential of the chiral unitary
 approach extending the predictions to higher energies than
 it would be possible with the use of $\chi PT$.
 Yet, we also saw that improvements in the basic
 information of the lowest order chiral Lagrangians to account for
 phenomenology of VMD are welcome.  On the other hand we found mandatory the
 inclusion of the $\pi\pi N $ channels in order to find an accurate
 reproduction of the data, particularly those in the isospin 3/2 sector.
 We also found that the introduction of these channels, forcing them to
 reproduce the inelasticities and other data, has as an indirect consequence
 that the $\Delta(1620)$ resonance appears then as a pole in the complex
 plane indicating a large mixing of this resonance with  $\pi\pi N $ states.
 This interpretation would be consistent with the large experimental coupling
 of this resonance to the $\pi\pi N $ channel.

\section*{Acknowledgements}

 We would like to acknowledge discussions with J.C. N\'acher.
 One of us, T. I., would like to thank J.A. Oller and A. Hosaka
 for useful discussions.
 This work has been partly supported by the Spanish Ministry of Education
 in the program 
 ``Estancias de Doctores y Tecn\'ologos Extranjeros en Espa\~na'',
 by the DGICYT contract number BFM2000-1326
 and by the EU TMR network Eurodaphne, contact no. ERBFMRX-CT98-0169.

\end{document}